\documentclass[12pt]{iopart}
\usepackage{graphicx}
\usepackage{float} 

\begin{document}

\title[Capturing rogue waves by multi-point statistics]{Capturing rogue waves by multi-point statistics}

\author{A. Hadjihosseini$^{1}$, Matthias W\"achter$^{1}$, N. P. Hoffmann$^{2,3}$ and J. Peinke$^{1,4}$}

\address{$^{1}$ Universit\"at Oldenburg, 26111 Oldenburg, Germany\\
$^{2}$ Hamburg University of Technology, 21073 Hamburg, Germany\\
$^{3}$ Imperial College, London SW7 2AZ, United Kingdom\\
$^{4}$ Fraunhofer Institute for Wind Energy and Energy System Technology, Ammerlander Heerstr. 136, DE-26129 Oldenburg, Germany}
\ead{\mailto{ali.hadjihosseini@uni-oldenburg.de}}
\begin{abstract}
As an example for complex systems with extreme events we investigate ocean wave states exhibiting rogue waves. We present a statistical method of data analysis based on multi-point statistics which for the first time allows grasping extreme rogue wave events in a statistically highly satisfactory manner. The key to the success of the approach is mapping the complexity of multi-point data onto the statistics of hierarchically ordered height increments for different time scales for which we can show that a stochastic cascade process with Markov properties is governed by a Fokker-Planck equation. Conditional probabilities as well as the Fokker-Planck equation itself can be estimated directly from the available observational data. With this stochastic description surrogate data sets can in turn be generated allowing to work out arbitrary statistical features of the complex sea state in general and extreme rogue wave events in particular. The results also open up new perspectives for forecasting the occurrence probability of extreme rogue wave events, and even for forecasting the occurrence of individual rogue waves based on precursory dynamics.
\end{abstract}

\pacs{
      02.50.Ga,  %Markov processes
      02.50.Ey,  %Stochastic processes
      05.45.Tp,  %Time series analysis
      89.75.-k ,   %Complex systems
      05.10.Gg,  %Stochastic analysis methods (Fokker-Planck, Langevin, etc.)
      92.10.Hm. %Ocean waves and oscillations
      }

%\keywords{Stochastic Process, Multi-point statistics, Rogue wave, Prediction}

%\submitto{\NJP}
\maketitle

\section{Introduction and motivation}
The observation and study of waves on the sea is probably one of the oldest scientific and cultural endeavours of mankind. But even today the sea's state can not be regarded as anything else than an enigma to man and science. Of course ocean waves have inspired a tremendous number of often groundbreaking results in mathematics, physics and related sciences, including nonlinear waves, localisation, extreme events, turbulence and many more. But still, the fully irregular and complex state of the sea is far from being understood. And both the characteristics of its irregularity, as well as the rare but extremely large wave events occurring sometimes, now often called rogue waves, abscond satisfying description, even in statistical terms.\\ 

Obviously the difficulties with understanding irregular and extreme or rogue ocean waves has to be seen in the context of extreme events in complex systems in general. Driven by various motives there has been extensive research on extreme events in a many fields, from the sciences, via meteorology and climate change, up to the social sciences and economics \cite{Palmer2002,Easterling2000,Onorato2013,Poon2004,Adlouni2008,Pisarenko2003}. It is still a strongly debated question whether extreme events are generally linked to some universal stochastic mechanisms or if they rather originate through special features of the individual systems under study  \cite{Sornette2009}. Still, a common point of all observations is that the empirical data are frequently punctuated by extreme events which seem to play an important role. Often an analysis approach is to approximate the observations by means of a generalized stochastic model in which some variables are represented in terms of stochastic components \cite{Chan2001}. Usually the complex systems under study are very high dimensional and thus finding adequate methods to model the stochastic components remains a challenge. 

Besides the description of extreme events in complex systems there is also the demand of their prediction.
Despite the fact that we have irregular and complex behavior of rogue waves, there are increasing number of research towards defining an early warning system for rogue wave occurrences \cite{Akhmediev2011,Akhmediev2011a} or establishing a prediction method for short term prediction of rogue events. Studies on  prediction methods mainly relys on deterministic behavior of non-stationary solutions of the underlying wave equations \cite{Olagnon2005,Alam2014,Islas2005} and also deterministic nonlinear time series analysis \cite{Birkholz2015}.\\

The present work is based on the finding that complex systems can often be described highly successfully as stochastic processes in scale rather than time or space. Examples are now known for various very different fields, like turbulence\cite{Friedrich1997,Renner2001}, economics\cite{Renner2001a,Nawroth2010,Ghasemi2007}, biology\cite{Friedrich2000,Ghasemi2006,Atyabi2006}, and many more, see \cite{Friedrich2011}. With scale dependent processes, originally introduced by Friedrich and Peinke \cite{Friedrich1997}, fractal and multi-fractal structures \cite{Friedrich2011} and even more generally joint n-point statistics \cite{Nawroth2010,Stresing2010} can be reduced by Markov properties to particular three point statistics. In a previous study \cite{Hadjihosseini2014} we have already shown for ocean wave data that certain scale dependent processes may have Markov properties. However, in \cite{Hadjihosseini2014} the Markov properties for the pure scale process could only be derived for deliberately pre-filtered data. In the present contribution we extend our previous investigations and base our analysis of the wave dynamics on general joint $(N+1)$-point probability density functions (PDF) $p(h(t),h(t-\tau_1), ... , h(t-\tau_N))$. Here $h(t)$ denotes the water surface elevation measured at a given location at time $t$ and $\tau_i$ are different time increments. The joint PDF provides the likelihood of a sequence of water surface elevation heights for $N+1$ different instants of time. We show how a Fokker-Planck equation can be derived which describes these general joint PDFs. Knowing the corresponding
stochastic process for the general multi-point statistics we can show that also the extreme events, i.e. the rogue waves, are grasped by this stochastic approach. The approach also allows time-series reconstruction in a statistical sense, and thus a statistically valid prediction of rogue wave occurrence.

The paper is structured as follows. First the mathematical aspects of multi-point and multi-scale description as well as the connection to scale dependent stochastic processes are introduced. Then the validity of the description based on observational ocean wave data is demonstrated. Finally the approach is applied to reconstruct time series for the underlying observational data and to forecast the occurrence probability of rogue waves in the given sea state.

%%%%%%%%%%%%%%%%%%%%%%%%%%%%%%%%%%%%%%%%%%%%%%%%%%%%%%%%%%
\section{Multi-point statistics} 
\label{multipoint}

In this section the statistical background of our approach for a multi-point reconstruction is presented. In the following we use the short-hand notation $h_i:=h(t_i)$ for the elevation of the water surface measured at a given location at time $t_i$, with $h_{i+j}:=h(t_i+t_j)$. We define the relative change in surface height over a time interval or, respectively, a time scale $\tau_j$ as
%%%%%%%%%%%% %%%%%%%%%%%%%%%%%%%%%%%         Eq 1            %%%%%%%%%%%%%%%%%%%%%%%%%%%%%%%%%
\begin{equation}
\xi_j\equiv \xi(\tau_j) :=h(t_i)-h(t_i-\tau_j) .
\label{inc}
\end{equation}
The aim is to calculate the joint probability $p(h^*,t^*;h_1,t^*-\tau_1;...;h_N,t^*-\tau_N)$ of occurrence of the event $\{ h^*,t^*\}$, together with the knowledge of  the past points $\{h_1,t^*-\tau_1;h_2,t^*-\tau_2;...;h_N,t^*-\tau_N\}$. We assume that the system has no explicit time dependence, i.e. the system is stationary. 
The probability of occurrence of the event $\{ h^*\}$ under the conditions of  the past points is given by

%%%%%%%%%%%% %%%%%%%%%%%%%%%%%%%%%%%         Eq 2            %%%%%%%%%%%%%%%%%%%%%%%%%%%%%%%%%

\begin{equation}
p(h^*|h_1,\tau_1;...;h_N,\tau_N)=\frac{p(h^* ; h_1,\tau_1 ; ... ; h_N,\tau_N)}{p(h_1,\tau_1 ; ... ; h_N,\tau_N)} .
\label{cpdf2}
\end{equation}

%%%%%%%%%%%% %%%%%%%%%%%%%%%%%%%%%%%%%%%%%%%%%%%%%%%%%%%%%%%%%%%%%%%%%%%%%%%%%
Next, the joint (N+1)-point PDF can be expressed in an equivalent way by joint increments statistics
%%%%%%%%%%%% %%%%%%%%%%%%%%%%%%%%%%%         Eq 3            %%%%%%%%%%%%%%%%%%%%%%%%%%%%%%%%%
\begin{eqnarray}
\nonumber
p(h^* ; h_1,\tau_1 ;h_2,\tau_2; ... ; h_N,\tau_N)&=&p(h^*-h_1,\tau_1;h^*-h_2,\tau_2;...;h^*-h_N,\tau_N,h^*)\\
\nonumber
&=&p(\xi_1;\xi_2;...;\xi_N,h^*)\\
&=&p(\xi_1;\xi_2;...;\xi_N|h^*)\cdot p(h^*) .
\label{npoint}
\end{eqnarray}

%%%%%%%%%%%% %%%%%%%%%%%%%%%%%%%%%%%%%%%%%%%%%%%%%%%%%%%%%%%%%%%%%%%%%%%%%%%%%
Note instead of the knowledge of wave heights at N+1 points, we consider now the knowledge of N height increments and one selected height $h^*$. Without loss of generality we take $\tau_i< \tau_{i+1}$, and thus introduce a hierarchical ordering of the increments $\xi_i$.

Only if the conditional PDFs do not depend on $h^*$, i. e. if 
%%%%%%%%%%%% %%%%%%%%%%%%%%%%%%%%%%%         Eq 4           %%%%%%%%%%%%%%%%%%%%%%%%%%%%%%%%%

\begin{equation}
p(\xi_1;\xi_2; ... ; \xi_N |h^*)=p(\xi_1;\xi_2; ... ;\xi_N), 
\label{indep1}
\end{equation}
%%%%%%%%%%%% %%%%%%%%%%%%%%%%%%%%%%%%%%%%%%%%%%%%%%%%%%%%%%%%%%%%%%%%%%%%%%%%%
the (N+1)-point statistics reduces to N-scale statistics of the increments $\xi_i$ at scales  $\tau_i$. In our previous work \cite{Hadjihosseini2014} we had applied filtering based on Hilbert-Huang transform techniques (HHT) to the wave data. The filtering removed the dependency on $h^*$ by kind of separating off the underlying dominant frequency, i.e. the wave like nature of the system. Still, already in this case Markov properties could be shown for the filtered wave data and thus the multi-scale PDF could be factorized in  
%%%%%%%%%%%% %%%%%%%%%%%%%%%%%%%%%%%         Eq 5           %%%%%%%%%%%%%%%%%%%%%%%%%%%%%%%%%

\begin{eqnarray}
\nonumber
p(h^* ; h_1,\tau_1 ; ... ; h_N,\tau_N)&=&p(\xi_1;...;\xi_N)\cdot p(h^*)\\
&=&p(\xi_1|\xi_2)\cdot...\cdot p(\xi_{N-1}|\xi_N)\cdot p(\xi_N)\cdot p(h^*).
\label{markov1}
\end{eqnarray}
%%%%%%%%%%%% %%%%%%%%%%%%%%%%%%%%%%%%%%%%%%%%%%%%%%%%%%%%%%%%%%%%%%%%%%%%%%%%%
In our present work here we do not apply any pre-filtering and stay focussed on the very direct data itself. This renders the approach much more general and we directly start with the multi-point statistics (Eq. (\ref{cpdf2}))
to investigate if the Markov property of the process is given for
%%%%%%%%%%%% %%%%%%%%%%%%%%%%%%%%%%%         Eq 6           %%%%%%%%%%%%%%%%%%%%%%%%%%%%%%%%%
\begin{equation}
p(\xi_j | \xi_{j+1},\xi_{j+2},...,\xi_{j+N},h^*)=p(\xi_j | \xi_{j+1},h^*).
\label{markov2}
\end{equation}
%%%%%%%%%%%% %%%%%%%%%%%%%%%%%%%%%%%%%%%%%%%%%%%%%%%%%%%%%%%%%%%%%%%%%%%%%%%%%
More specifically we will first investigate from the observational data if
%%%%%%%%%%%% %%%%%%%%%%%%%%%%%%%%%%%         Eq 7           %%%%%%%%%%%%%%%%%%%%%%%%%%%%%%%%%
\begin{equation}
p(\xi_j | \xi_{j+1},\xi_{j+2},h^*)=p(\xi_j | \xi_{j+1},h^*)
\label{markov3}
\end{equation}
%%%%%%%%%%%% %%%%%%%%%%%%%%%%%%%%%%%%%%%%%%%%%%%%%%%%%%%%%%%%%%%%%%%%%%%%%%%%%
holds. This we will take as hint for the validity of Markov property. Using Eq. (\ref{markov2}), the multi-point PDF Eq. (\ref{npoint}) can then be factorised as 

%%%%%%%%%%%% %%%%%%%%%%%%%%%%%%%%%%%         Eq 8          %%%%%%%%%%%%%%%%%%%%%%%%%%%%%%%%%
\begin{equation}
p(h^* ; h_1,\tau_1 ;h_2,\tau_2; ... ; h_N,\tau_N)=p(\xi_1|\xi_2,h^*)\cdot...\cdot p(\xi_{N-1}|\xi_N,h^*)\cdot p(\xi_N|h^*)\cdot p(h^*) .
\label{npoint2}
\end{equation}
%%%%%%%%%%%% %%%%%%%%%%%%%%%%%%%%%%%%%%%%%%%%%%%%%%%%%%%%%%%%%%%%%%%%%%%%%%%%%
As Eq. ( \ref{markov2}) is nothing else than the Markov property of a stochastic process of $\xi_i$ evolving in
the time scale $\tau_i$,
the evolution of conditional PDFs of Eq. (\ref{npoint2}) can be expressed by Kramers-Moyal expansion \cite{Risken1989},

%%%%%%%%%%%% %%%%%%%%%%%%%%%%%%%%%%%         Eq 9         %%%%%%%%%%%%%%%%%%%%%%%%%%%%%%%%%
\begin{equation}
-\tau_j\frac{\partial}{\partial\tau_j}p(\xi_j|\xi_k,h^*)=\sum_{n=1}^{\infty}(-\frac{\partial}{\partial \xi_j})^n\left[ D^{(n)}(\xi_j,\tau_j,h^*)p(\xi_j|\xi_k,h^*)\right], 
\label{km}
\end{equation}
%%%%%%%%%%%% %%%%%%%%%%%%%%%%%%%%%%%%%%%%%%%%%%%%%%%%%%%%%%%%%%%%%%%%%%%%%%%%%
where Kramers-Moyal coefficients $D^{(n)}$ are defined as 
%%%%%%%%%%%% %%%%%%%%%%%%%%%%%%%%%%%         Eq 10          %%%%%%%%%%%%%%%%%%%%%%%%%%%%%%%%%
\begin{equation}
D^{(n)}(\xi_j,\tau_j,h^*)=\lim_{\delta_\tau \to 0} \frac{\tau_j}{n!\delta\tau}<\left[\xi_j^{\prime}(\tau_j-\delta\tau,h^*)-\xi_j(\tau_j,h^*)\right]^n>_{\xi_j^{\prime}}.
\label{km_coef}
\end{equation}
%%%%%%%%%%%% %%%%%%%%%%%%%%%%%%%%%%%%%%%%%%%%%%%%%%%%%%%%%%%%%%%%%%%%%%%%%%%%%
Note that the pre-factor $- \tau$ in Eq. (\ref{km}) indicates that we consider the process for decreasing $\tau$- values and an evolution in log-scale of $\tau$.

If the Kramers-Moyal coefficient $D^{(4)}$ is zero, then it follows from the Pawula theorem that all coefficients for $n \geq 3$ are zero, too cf. \cite{Risken1989}. The Kramers-Moyal expansion then yields a Fokker-Planck equation with just two coefficients,

%%%%%%%%%%%% %%%%%%%%%%%%%%%%%%%%%%%         Eq 11          %%%%%%%%%%%%%%%%%%%%%%%%%%%%%%%%%
\begin{equation}
-\tau_j\frac{\partial}{\partial\tau_j}p(\xi_j|\xi_k,h^*)=-\frac{\partial}{\partial\xi_j}\left[D^{(1)}(\xi_j,\tau_j,h^*)p(\xi_j|\xi_k,h^*)\right]\\
+\frac{\partial^2}{\partial\xi_j^2}\left[D^{(2)}(\xi_j,\tau_j,h^*)p(\xi_j|\xi_k,h^*)\right].
\label{fp}
\end{equation}
%%%%%%%%%%%% %%%%%%%%%%%%%%%%%%%%%%%%%%%%%%%%%%%%%%%%%%%%%%%%%%%%%%%%%%%%%%%%%
$D^{(1)}$ denotes the drift and $D^{(2)}$ the diffusion coefficient. With this the Fokker-Planck equation turns out a suitable description for the conditional probabilities of the water surface height increments, from which in turn the general multi-point joint PDF of the surface heights themselves, Eq. (\ref{npoint2}), can be determined as

%%%%%%%%%%%% %%%%%%%%%%%%%%%%%%%%%%%         Eq 12          %%%%%%%%%%%%%%%%%%%%%%%%%%%%%%%%%
\begin{eqnarray}
\nonumber
p(h^* | h_1,\tau_1 ; ... ; h_N,\tau_N)&=\frac{p(h^*)}{p(h_1,\tau_1)}\times p(h^*-h_1,\tau_1|h^*-h_2,\tau_2;h^*)\times&\\
\nonumber
&\times ... \times \frac{p(h^*-h_{N-1},\tau_{N-1}|h^*-h_N,\tau_N;h^*)}{p(h_1-h_{N-1},\tau_{N-1}-\tau_1|h_1-h_N,\tau_N-\tau_1;h_1,\tau_1)}&\\
\nonumber
&\times \frac{p(h^*-h_N,\tau_N|h^*)}{p(h_1-h_N,\tau_N-\tau_1|h_1,\tau_1)}.&
\label{mp}
\end{eqnarray}

%%%%%%%%%%%% %%%%%%%%%%%%%%%%%%%%%%%%%%%%%%%%%%%%%%%%%%%%%%%%%%%%%%%%%%%%%%%%%
Using the increment notation, omitting the $\tau$-values and 
defining $\tilde{\xi}_j := h_1-h_j$ with the corresponding time scale $\tau_j-\tau_1$ and $j=2, \dots, N$, this equation simplifies to 

%%%%%%%%%%%% %%%%%%%%%%%%%%%%%%%%%%%         Eq 12          %%%%%%%%%%%%%%%%%%%%%%%%%%%%%%%%%
%\begin{equation}
%\begin{aligned}
%&p(h^* | h_1,\tau_1 ; ... ; h_N,\tau_N)=&\\
%&\frac{p(h^*)}{p(h_1)}\times p(\xi_1|\xi_2;h^*)\times ... \times \frac{p(\xi_{N-1}|\xi_N;h^*)}{p(\tilde{\xi}_{N-1},|\tilde{\xi}_N;h1)}\times \frac{p(\xi_N|h^*)}{p(\tilde{\xi}_N|h_1)}.&
%\label{mp2}
%\end{aligned}
%\end{equation}
\begin{equation}
p(h^* | h_1,\tau_1 ; ... ; h_N,\tau_N)=\frac{p(h^*)}{p(h_1)}\times \frac{\prod_{i=1}^{N-1} p(\xi_{i}|\xi_{i+1};h^*)}{\prod_{i=2}^{N-1} p(\tilde{\xi}_{i},|\tilde{\xi}_{i+1};h1)}\times \frac{p(\xi_N|h^*)}{p(\tilde{\xi}_N|h_1)}.
\label{mp2}
\end{equation}
%%%%%%%%%%%% %%%%%%%%%%%%%%%%%%%%%%%%%%%%%%%%%%%%%%%%%%%%%
%%%%%%%%%%%%%%%%%%%%
For a given height $h^*$ the probability of its occurrence $p(h^* | h_1,\tau_1 ; ... ; h_N,\tau_N)$ is given by the simple conditional PDFs, which can be calculated from the Fokker-Planck equation, or which can be estimated directly from the data. Note the simple conditional PDFs $p(\xi_i,\tau_i|\xi_j,\tau_j;h^*)$ only contain information about three height values $h^*,h_i,h_j$; or more abstractly, of three points of the time series $h(t)$. Thus Eq. (\ref{mp}) is a three point closure of the multi-point problem.

%%%%%%%%%%%%%%%%%%%%%%%%%%%%%%%           Experimental results     %%%%%%%%%%%%%%%%%%%%%%%%%%%%%%%
\section{Results based on observational data}
\label{experimentalresults}
\begin{figure}[thb]
  \begin{center}
    \includegraphics[width=.48\textwidth]{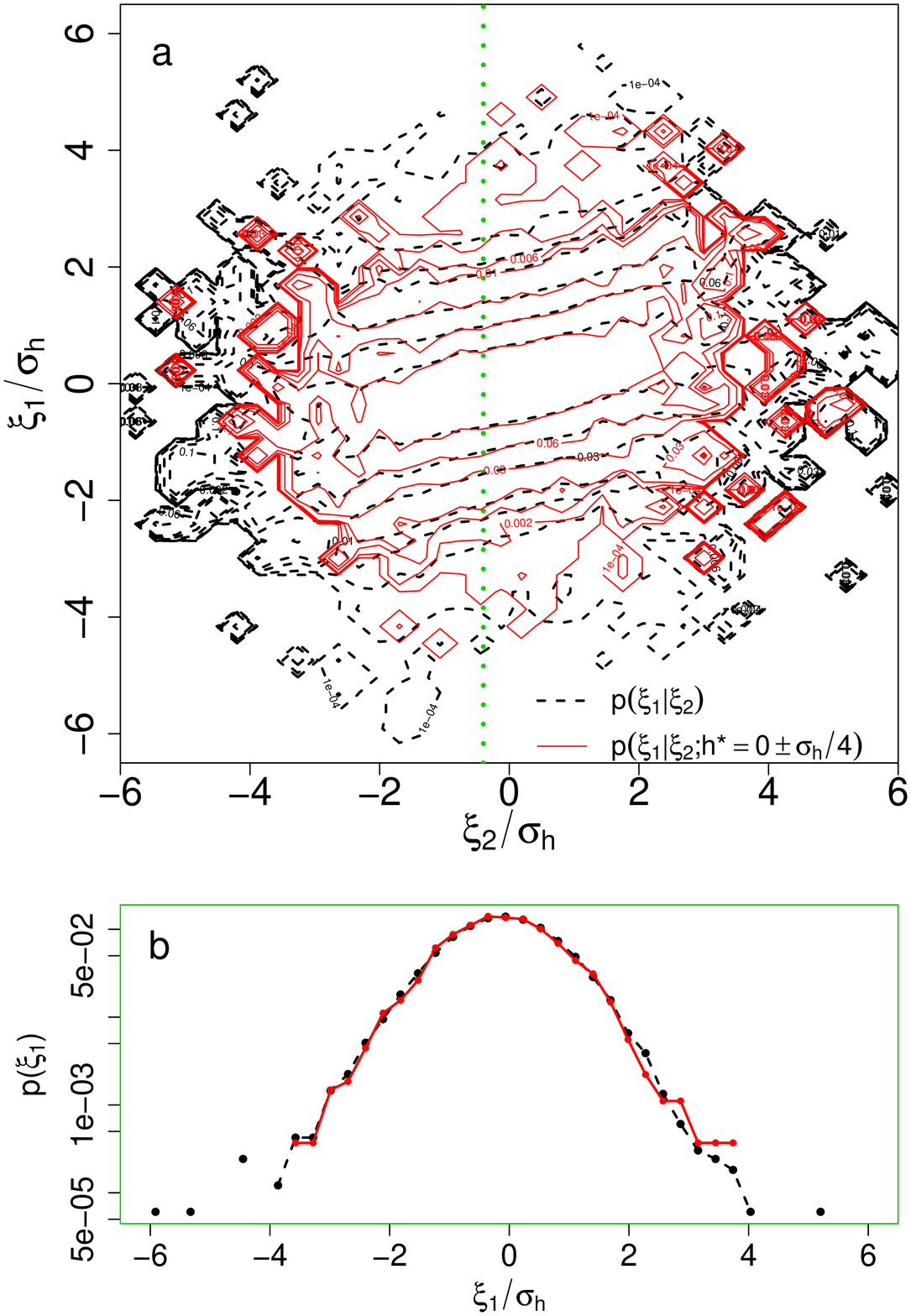}
    \includegraphics[width=.48\textwidth]{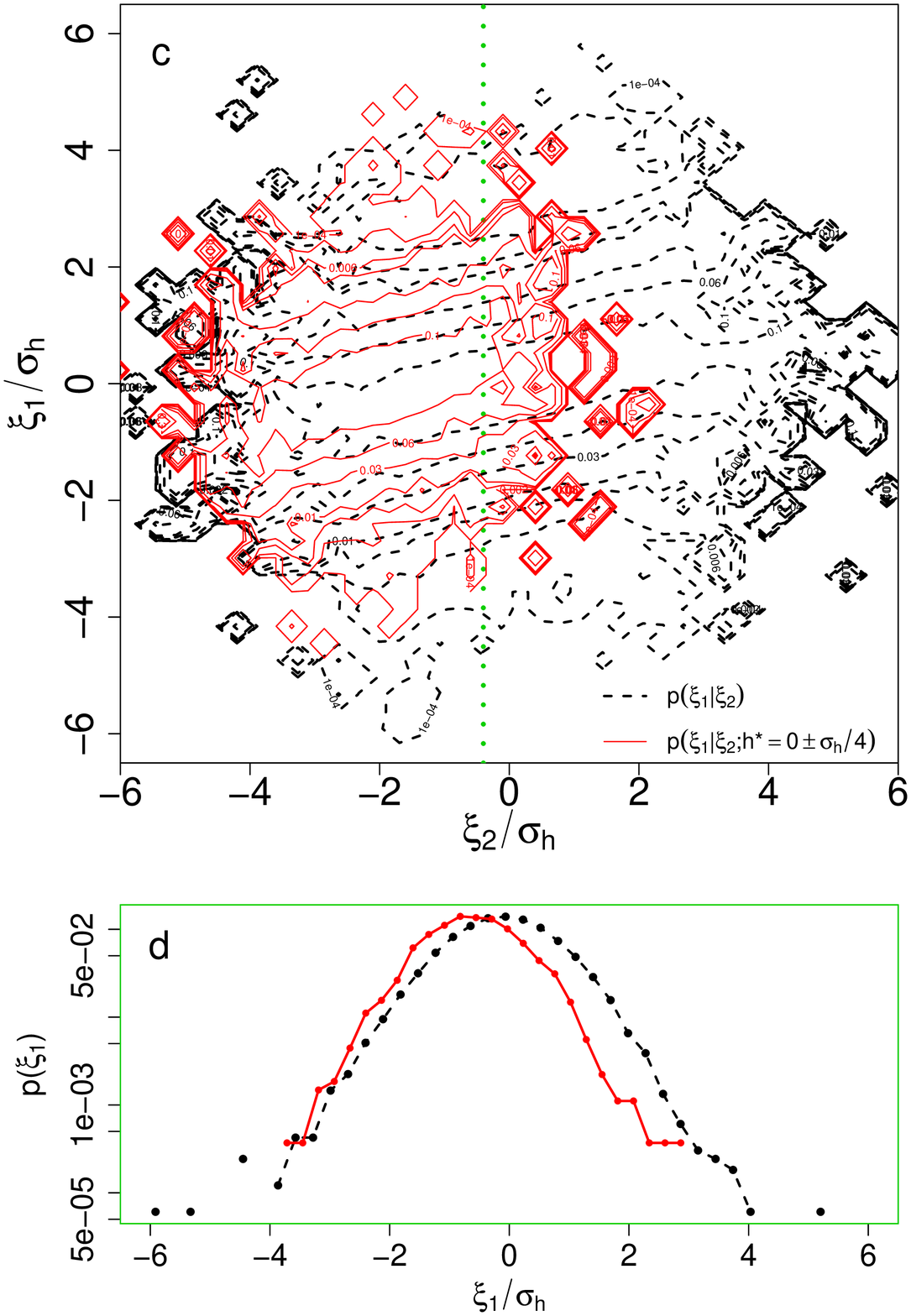}
  \end{center}
 \caption{Contour plots of the conditional PDFs $p(\xi_1|\xi_2)$ (dashed black lines) and $p(\xi_1 | \xi_2; h^* )$ (solid red lines) with $\xi_j \equiv \xi(\tau_j)$ for scales $\tau_1=14$ and $\tau_2=2\tau_1$, for $h^*=0\pm \sigma_h/4$ (a) and $h^*=2\sigma_h \pm \sigma_h/4$ (c). Cuts through the conditional PDFs for fixed values of $\xi_2 = -0.5\sigma_h$ in (a) and (c) are shown in (b) and (d) respectively.}
 \label{cond1to2}
\end{figure}

The wave measurements used in this study were taken in the Sea of Japan, at a location 3 km off the Yura fishery harbor, where the water depth is about 43 meters, further details can be found in  \cite{Mori2002,Mori2002a,Mori2002b,Mori2000}. 
First we want to examine if the conditional PDFs depend on the wave height itself by comparing both sides of the equation
%%%%%%%%%%%% %%%%%%%%%%%%%%%%%%%%%%%         Eq 13         %%%%%%%%%%%%%%%%%%%%%%%%%%%%%%%%%
\begin{equation}
p(\xi_1|\xi_2; h^*)=p(\xi_1|\xi_2).
\label{pdf}
\end{equation}
%%%%%%%%%%%% %%%%%%%%%%%%%%%%%%%%%%%%%%%%%%%%%%%%%%%%%%%%%%%%%%%%%%%%%%%%%%%%%
In Fig. \ref{cond1to2} the comparison of conditional PDFs from both sides of Eq. (\ref{pdf}) at scales $\tau_1=14$ and $\tau_2=28$ seconds and for two different values of $h^*$ are shown. To get sufficient data we always use an interval of $h^*$ with $\pm \sigma_h/4$ (where $\sigma_h=\sqrt{\langle h^2 \rangle}$ ).
For $h^*=0$ in  Fig. \ref{cond1to2}(a) both distributions are almost the same but for values of $h_0 \ne 0$, like in Fig. \ref{cond1to2}(c)  a significant shift of the red contour plot (solid lines), which is the left hand side of Eq. (\ref{pdf}), is found. As .epsa result from this one can clearly deduce that the conditional PDFs $p(\xi_1|\xi_2; h^*)$ do depend on $h^*$. \\

%%%%%%%%%%%% %%%%%%%%%%%%%%%%%%%%%%%         Fig.1          %%%%%%%%%%%%%%%%%%%%%%%%%%%%%%%%%
\begin{figure}[thb]
  \begin{center}
    \includegraphics[width=.96\textwidth]{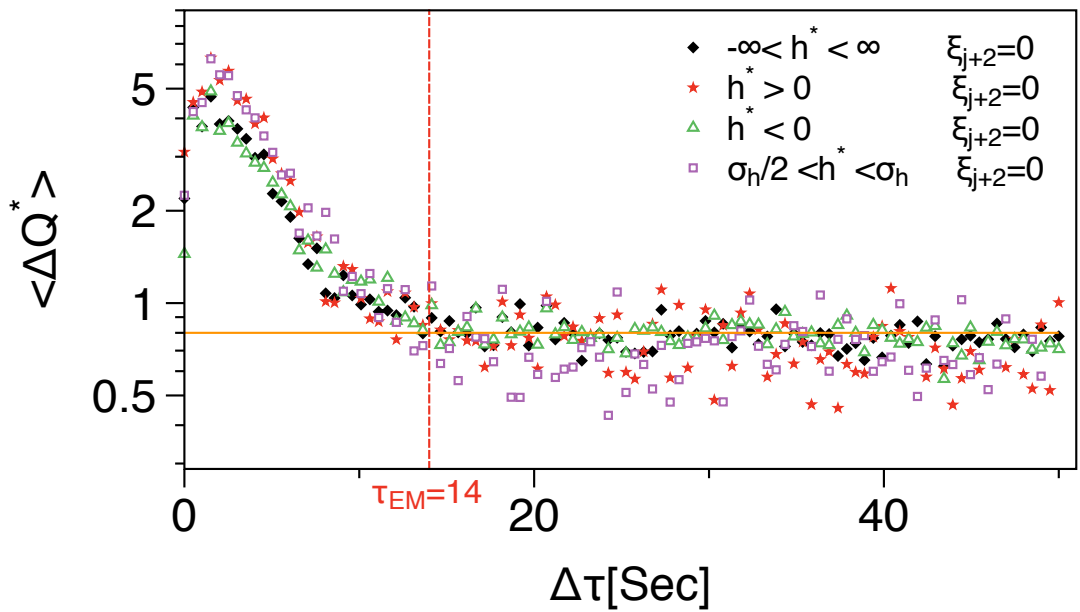}
   \end{center}
 \caption{Wilcoxon test of Eq. \ref{markov3} for different values of $h^*$.}
 \label{wilcoxon}
\end{figure}

%%%%%%%%%%%% %%%%%%%%%%%%%%%%%%%%%%%%%%%%%%%%%%%%%%%%%%%%%%%%%%%%%%%%%%%%%%%%%
Next the Markov properties according to Eq. (\ref{markov3}) can be checked. Note that we have to compare
two data sets according to $\xi_j|_{\xi_{j+1};h^*}$ and $\xi_j|_{\xi_{j+1};\xi_{j+2};h^*}$, and the size of each of these data sets is very different. The verification is thus performed by the use of the Wilcoxon test \cite{Renner2001} as this test is suitable to compare the statistical similarity of two sample sets of different sizes.  The validity of the Wilcoxon test can be shown by the
normalized expectation value $<\Delta Q^*>$ of the number of inversions of the conditional wave height increments $\xi_j|_{\xi_{j+1};h^*}$ and $\xi_j|_{\xi_{j+1};\xi_{j+2};h^*}$. If Markov properties are given, $<\Delta Q^*>$ has a value of $\sqrt{2/\pi}\approx 0.8$. The values of $<\Delta Q^*>$ in Fig. \ref{wilcoxon} for different values of $h^*$ show that Markov properties hold for  $(\tau \ge14$ seconds$)$. This defines a finite minimum step size or scale in the Markov process of the evolution of the surface elevation increments $\xi_i$.Such a finite step size is well known for stochastic processes in general \cite{Einstein1905}, and the scale is called the Einstein-Markov length, which has for example also been found in a similar way for turbulent flow data, cf. \cite{Lueck2006,Friedrich1998}. The scale has been marked by a vertical red dashed line in Fig. \ref{wilcoxon}. Note that compared to our previous work \cite{Hadjihosseini2014} the Markov properties are fulfilled without applying a Hilbert-Huang Transform (HHT) to the original data, which is due to the fact that we have now included the dependencies on $h^*$.

%%%%%%%%%%%% %%%%%%%%%%%%%%%%%%%%%%%         Fig.2          %%%%%%%%%%%%%%%%%%%%%%%%%%%%%%%%%
\begin{figure}[H]
\begin{center}
  \includegraphics[width=0.8\textwidth]{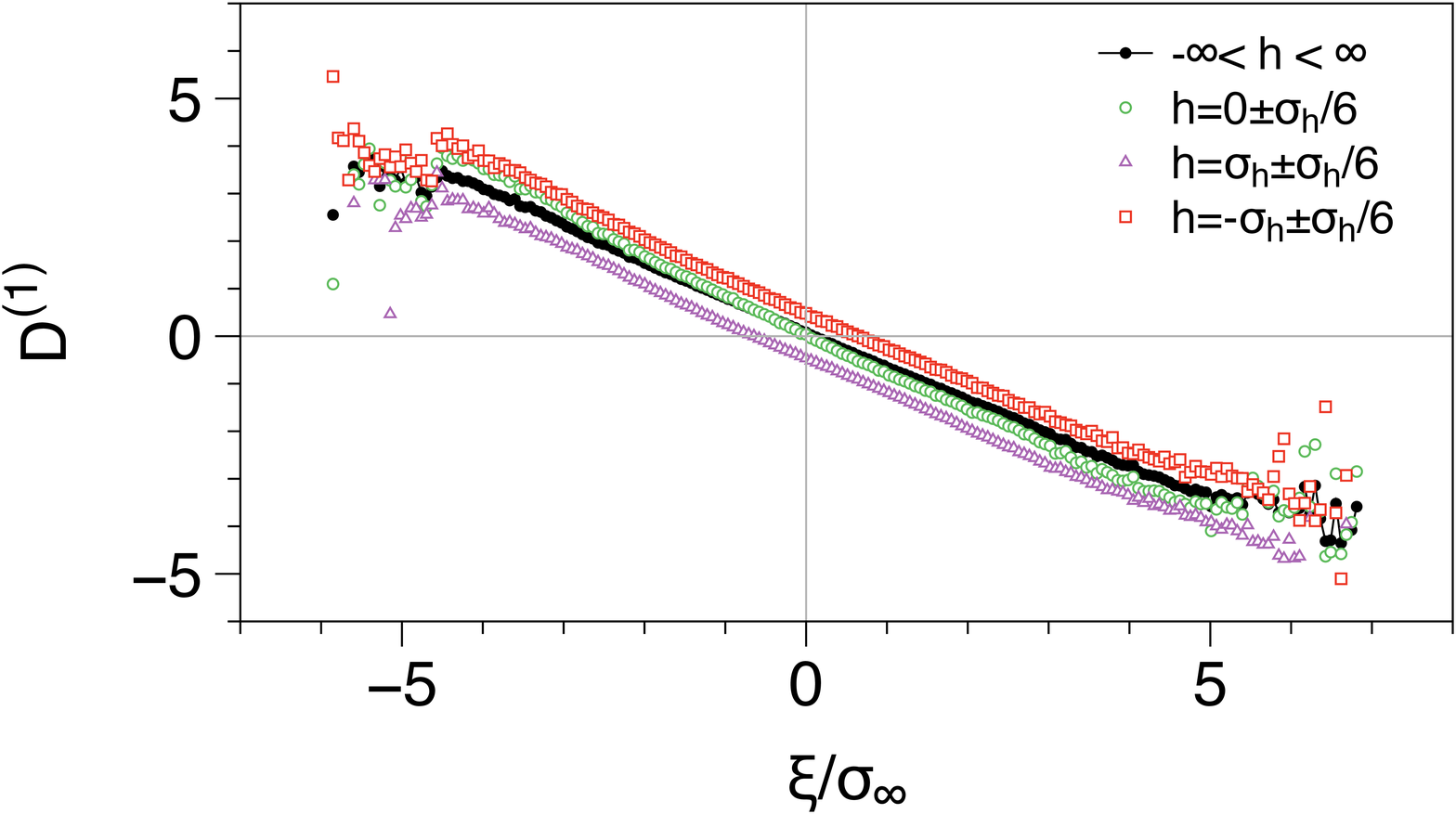}
    \includegraphics[width=0.8\textwidth]{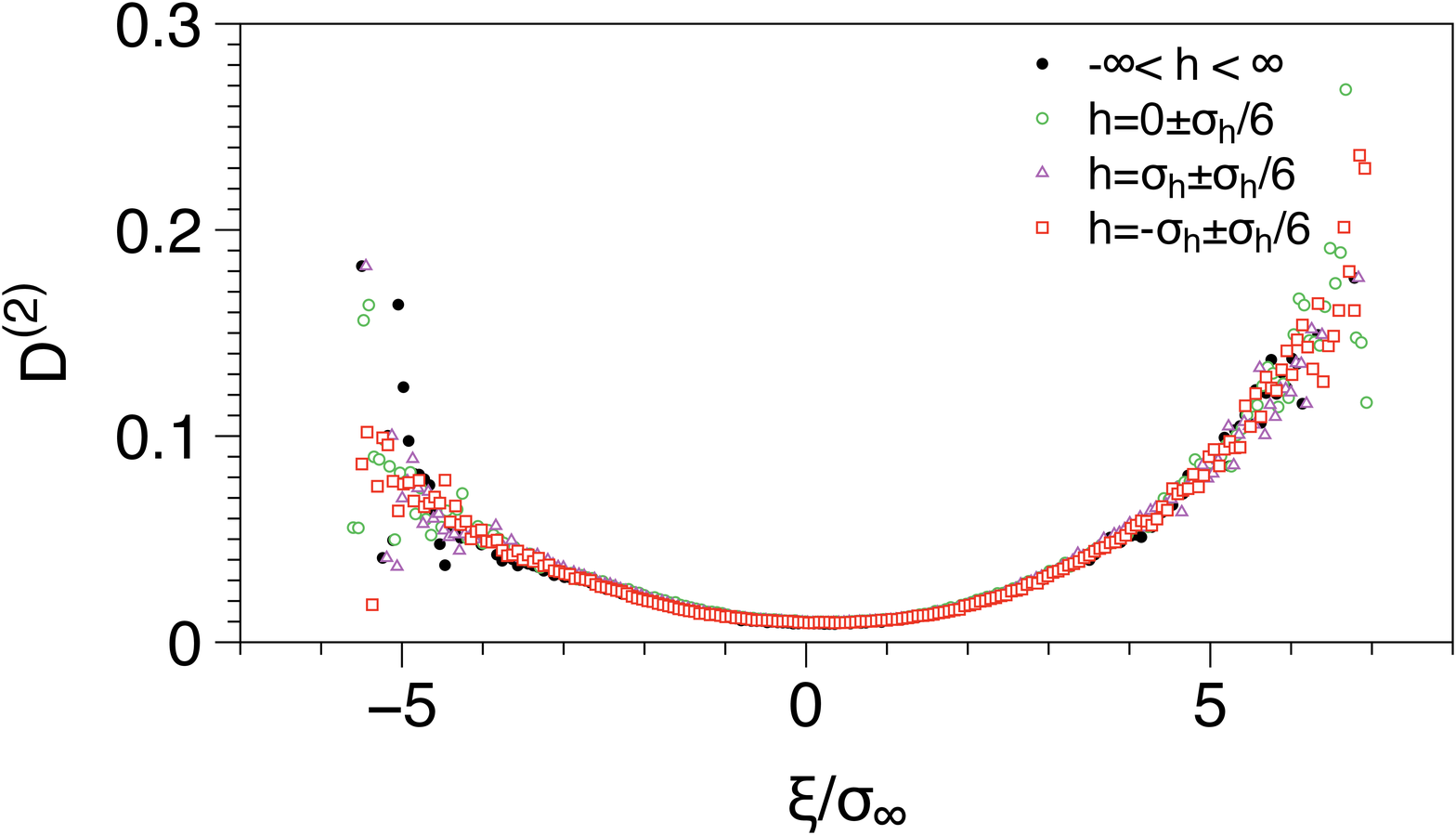}
        \includegraphics[width=0.8\textwidth]{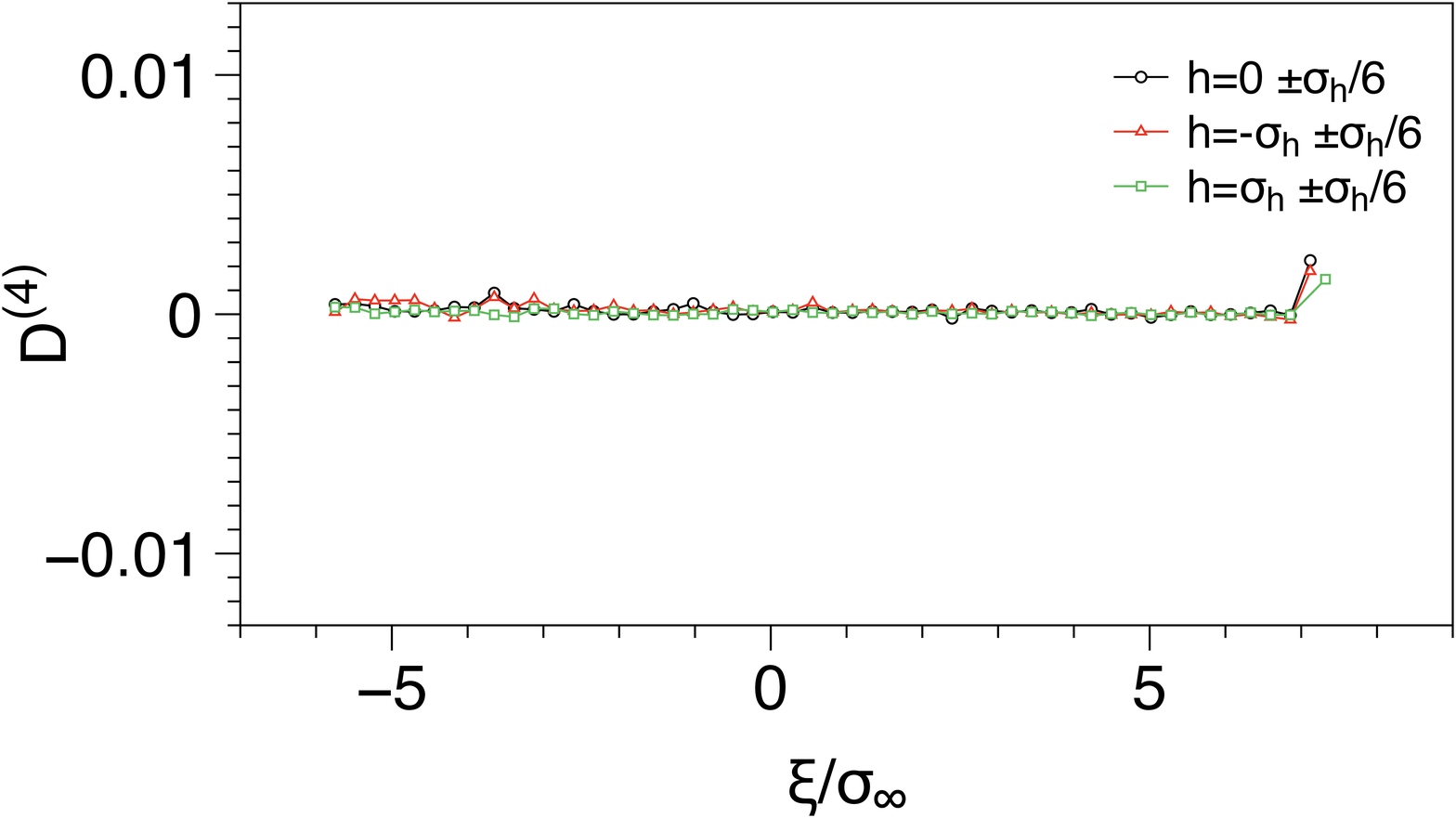}
    \end{center}
    \caption{Drift, $D^{(1)}(\xi,\tau,h)$, diffusion, $D^{(2)}(\xi,\tau,h)$, and the Kramers-Moyal coefficient $D^{(4)}(\xi,\tau,h)$ for different value of wave height, $h$, at $\tau=140$ $s$.}
    \label{d1d2}
  \end{figure}

%%%%%%%%%%%% %%%%%%%%%%%%%%%%%%%%%%%%%%%%%%%%%%%%%%%%%%%%%%%%%%%%%%%%%%%%%%%%%
Based on the finding that Markov properties are fulfilled for the evolution of water surface height increments $\xi_j$ with decreasing time scale $\tau_j$ we can now proceed to estimate the corresponding stochastic process via the above mentioned Kramers-Moyal coefficients. Based on the knowledge of the conditional PDF like shown in Fig. \ref{cond1to2} the conditional average in Eq. (\ref{km_coef}) is known too. The estimation of $\lim_{\delta_\tau \to 0}$ causes some problems, in particular due to the Einstein-Markov length, but has been worked out already in several publications \cite{Renner2001,Boettcher2006,Gottschall2008}. Besides this direct estimation we optimize the obtained functional forms of  $D^{(1)}$ and $D^{(2)}$ by minimizing the differences between measured conditional PDFs and those obtained by numerical solutions of the resulting Fokker-Planck equation \cite{Nawroth2007}. 
Fig. \ref{d1d2} shows the estimated drift and diffusion functions, $D^{(1)}(\xi,\tau,h)$ and $D^{(2)}(\xi,\tau,h)$, of ocean wave surface elevation data for $\tau=140$ seconds and different values of wave height $h^*$. For $h^*\ne 0$ the $D^{(1)}(\xi,\tau,h)$ curves are shifted in the vertical direction, whereas no significant change is found for the diffusion term. Furthermore the fourth order Kramers-Moyal coefficient $D^{(4)}(\xi,\tau,h)$ is indeed found to be close to zero for different values of $h^*$, thus the Fokker-Planck description we propose in Eq. (\ref{fp}) can be assumed to be valid.  Note that the surface elevation height increments, $\xi$, are given in the units of their standard deviation in the limit $\tau \to \infty$, $\sigma_\infty$, which is identical to $\sqrt{2}\sigma_h\equiv \sqrt{2\langle h^2\rangle}$ \cite{Renner2001}.

%%%%%%%%%%%% %%%%%%%%%%%%%%%%%%%%%%%         FIG 3            %%%%%%%%%%%%%%%%%%%%%%%%%%%%%%%%%

\begin{figure}[thb]
  \begin{center}
    \includegraphics[width=.9\textwidth]{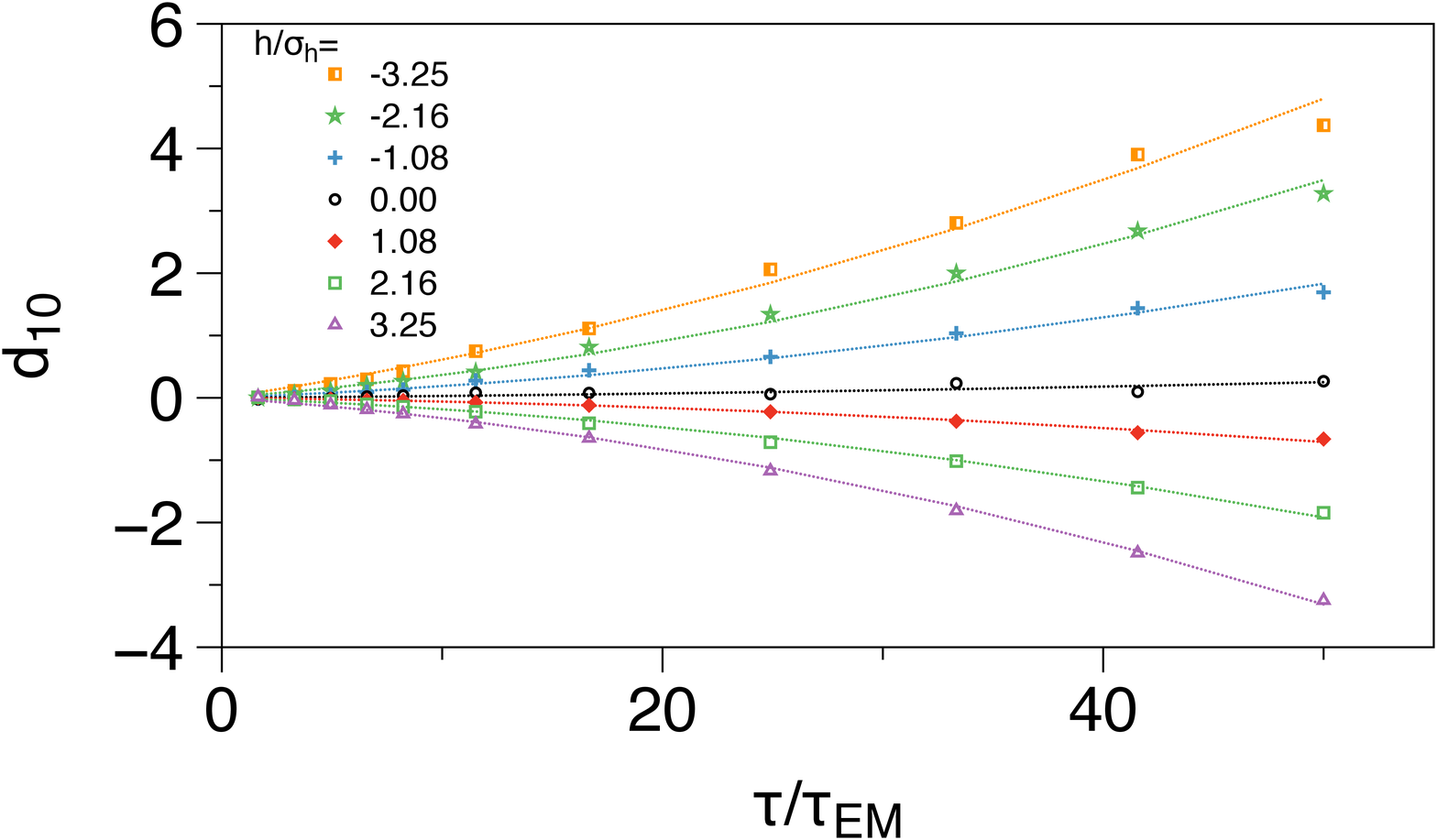}
   \end{center}
 \caption{coefficient $d_{10}$ from Eq. \ref{d100} as a function of $\tau$ for different wave surface height values, $h^*$. The dotted lines are the second-order polynomial fits in $\tau$.}
 \label{d10}
\end{figure}

%%%%%%%%%%%% %%%%%%%%%%%%%%%%%%%%%%%%%%%%%%%%%%%%%%%%%%%%%%%%%%%%%%%%%%%%%%%%%
To ease parameterisation, the drift and diffusion terms can be approximated by first and second order polynomials in $\xi$,
%%%%%%%%%%%% %%%%%%%%%%%%%%%%%%%%%%%         Eq 14          %%%%%%%%%%%%%%%%%%%%%%%%%%%%%%%%%
\begin{eqnarray}
	\nonumber
	D^{(1)}(\xi,\tau,h) &=& d_{10}(\tau,h) -d_{11}(\tau)\xi,\\
	D^{(2)}(\xi,\tau) &=& d_{20}(\tau)-d_{21}(\tau)\xi+d_{21}(\tau)\xi^2.
	\label{d100}
	\end{eqnarray}
%%%%%%%%%%%% %%%%%%%%%%%%%%%%%%%%%%%%%%%%%%%%%%%%%%%%%%%%%%%%%%%%%%%%%%%%%%%%%
The height dependency of the drift function is expressed by the $d_{10}(h^*,\tau)$-coefficient and our results are shown in Fig. \ref{d10}.  The results indicate once more that we have a strong wave height dependency in our process. \\ 
%%%%%%%%%%%% %%%%%%%%%%%%%%%%%%%%%%%         Fig.4          %%%%%%%%%%%%%%%%%%%%%%%%%%%%%%%%%

\section{Reconstruction of time series}

\begin{figure}[thb]
\begin{center}
    \includegraphics[width=0.9\textwidth]{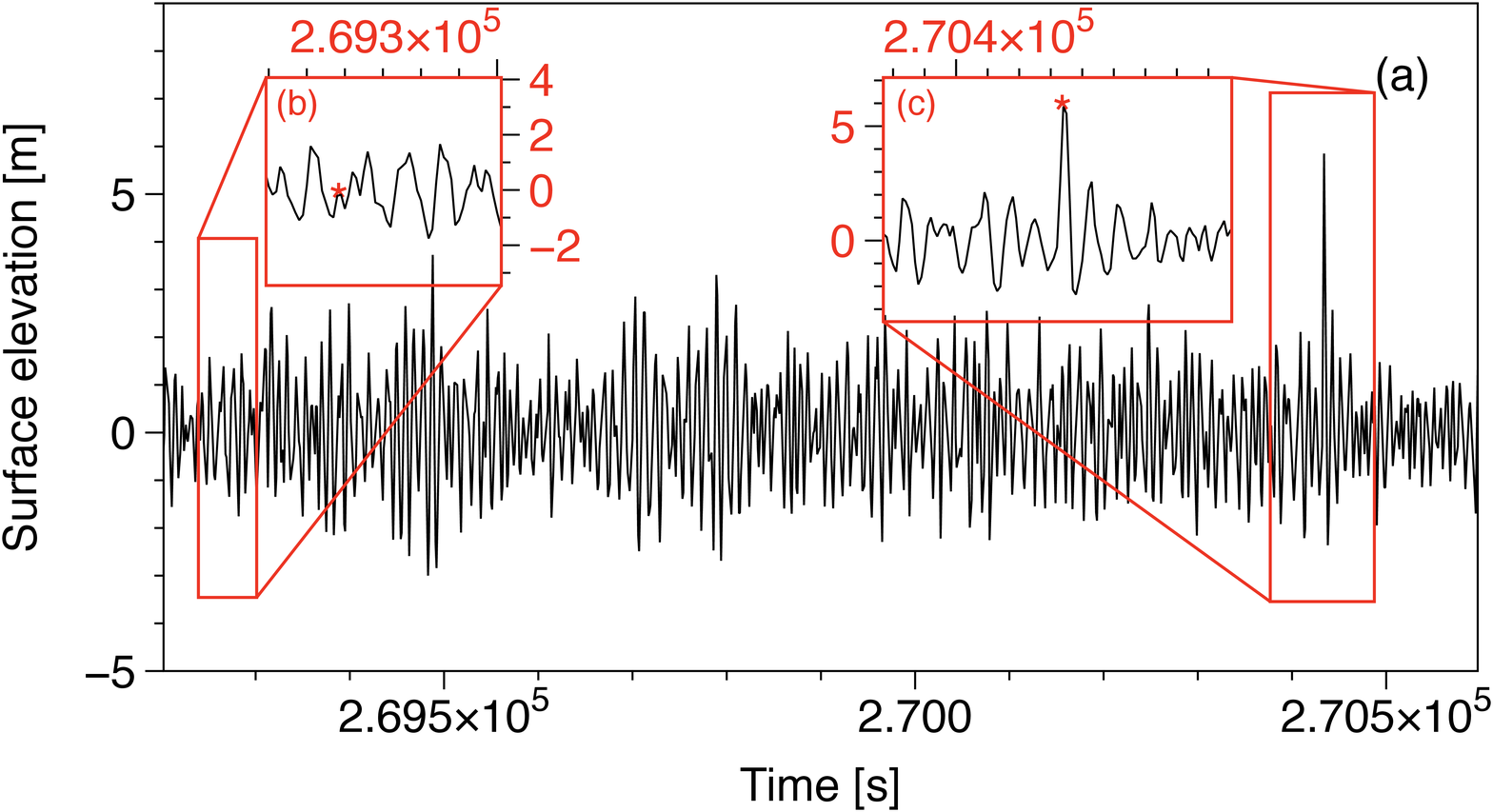}
     \includegraphics[width=0.43\textwidth]{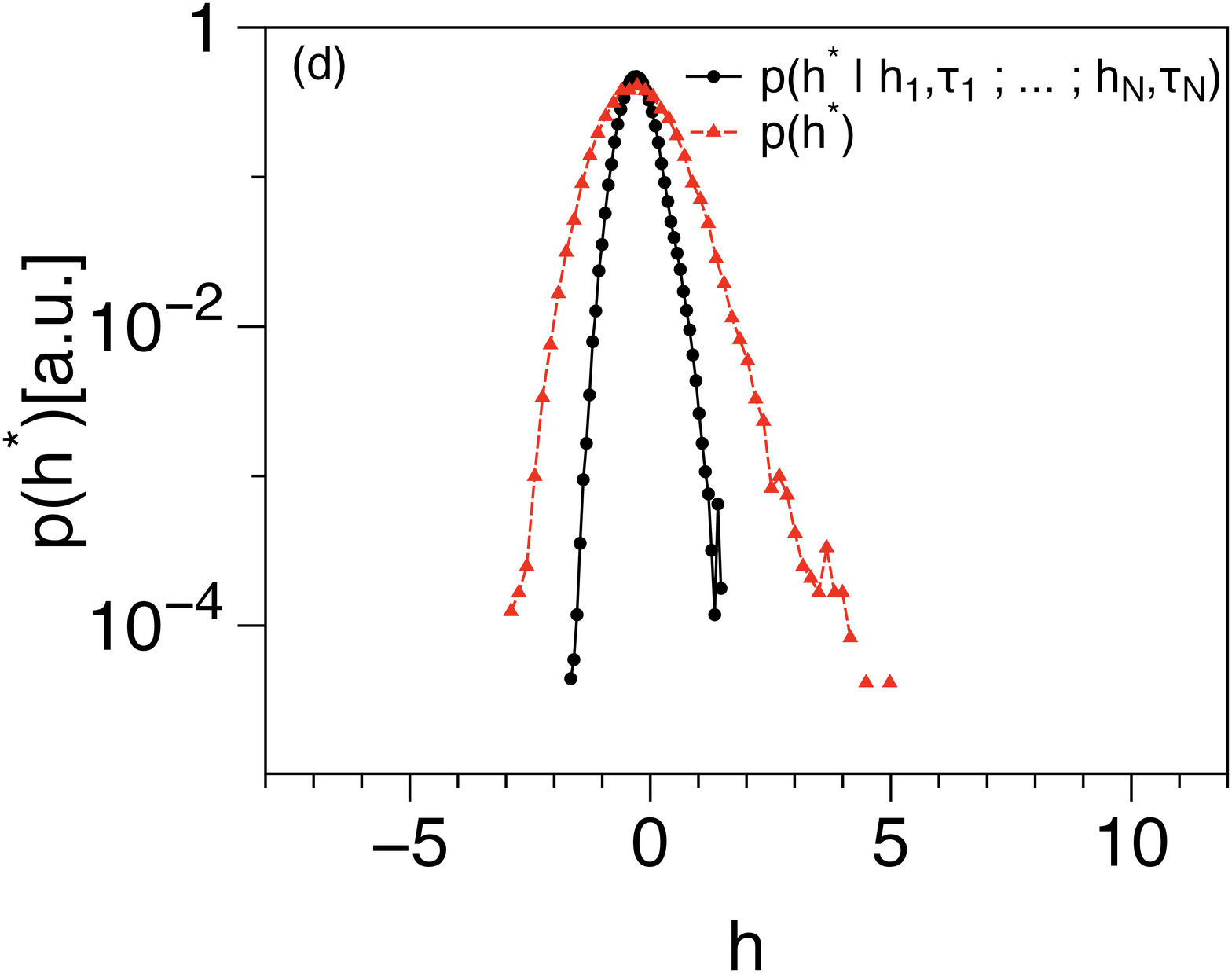}
      \includegraphics[width=0.43\textwidth]{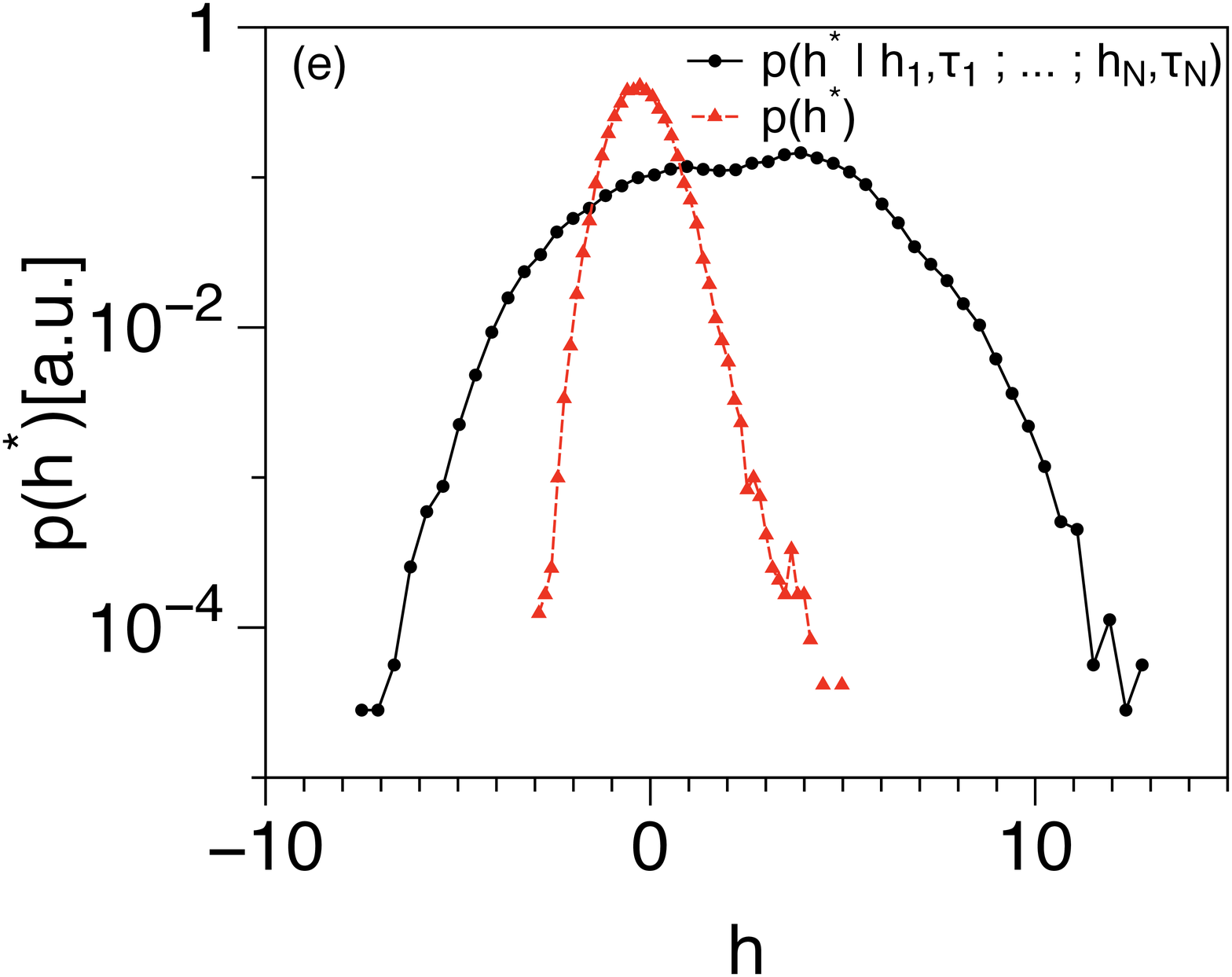}
    
    \end{center}
    \caption{Reconstructed time series (a) after Eq. \ref{mp2}. Two time windows are marked by $(b)$ and $(c)$  for which the corresponding multi- conditioned PDFs are given (d) and (e). To show the changing volatility the multi- conditioned PDFs  (black), the unconditional PDFs  (red) estimation from all data are shown too. Note the obvious changes of the likelihood of large wave amplitudes. }
    \label{recon_time_series}
  \end{figure}

The knowledge of the conditional probabilities $p(h^*|h_1,\tau_1, \dots, h_N,\tau_N)$ and its estimation by Eq. (\ref{mp2}) can be used to generate a new data point $h^*$. Shifting the procedure by one step and repeating the same procedure may be used to generate new surrogate time series.  For technical reasons one should avoid zeros in conditional pdfs if one uses Eq. (\ref{mp2}). Here we used kernel density estimation which is very helpful for parameter ranges for which we have only limited data \cite{Green1993,Hastie1990}. The initial idea for reconstructing time-series following this procedure was originally developed in a similar way for fluid turbulence data, see \cite{Nawroth2006}. The time scales we use here for this process are $\tau_n=n\cdot \tau_{EM}$ where $n=1,2,\dots,7$ and the Einstein-Markov time scale $\tau_{EM}=14$ seconds, as shown in Fig. \ref{wilcoxon}. (The maximal value of $n=7$ was chosen, as for that time step the autocorrelation of the height increments approaches zero.) In Fig. \ref{recon_time_series}(a) typical time series obtained is shown. In the figure part (d) and (e) two selected conditional probabilities $p(h^*|h_1,\tau_1, \dots, h_N,\tau_N)$ are shown to illustrate our method. In addition to the conditional probabilities the single event probability $p(h^*)= p(h)$ of all height values is shown (red curve). These figures show clearly how the conditional probabilities change with $h_1,\tau_1, \dots, h_N,\tau_N$ the values of the N wave heights seen before. There are cases when smaller $h^*$- values are expected in the next step, see Fig. \ref{recon_time_series} (b), and there are cases when large $h^*$- values become highly likely, see Fig. \ref{recon_time_series} (c).  
 
%%%%%%%%%%%% %%%%%%%%%%%%%%%%%%%%%%%         FIG 5           %%%%%%%%%%%%%%%%%%%%%%%%%%%%%%%%%
\begin{figure}[thb]
  \begin{center}
    \includegraphics[width=.47\textwidth]{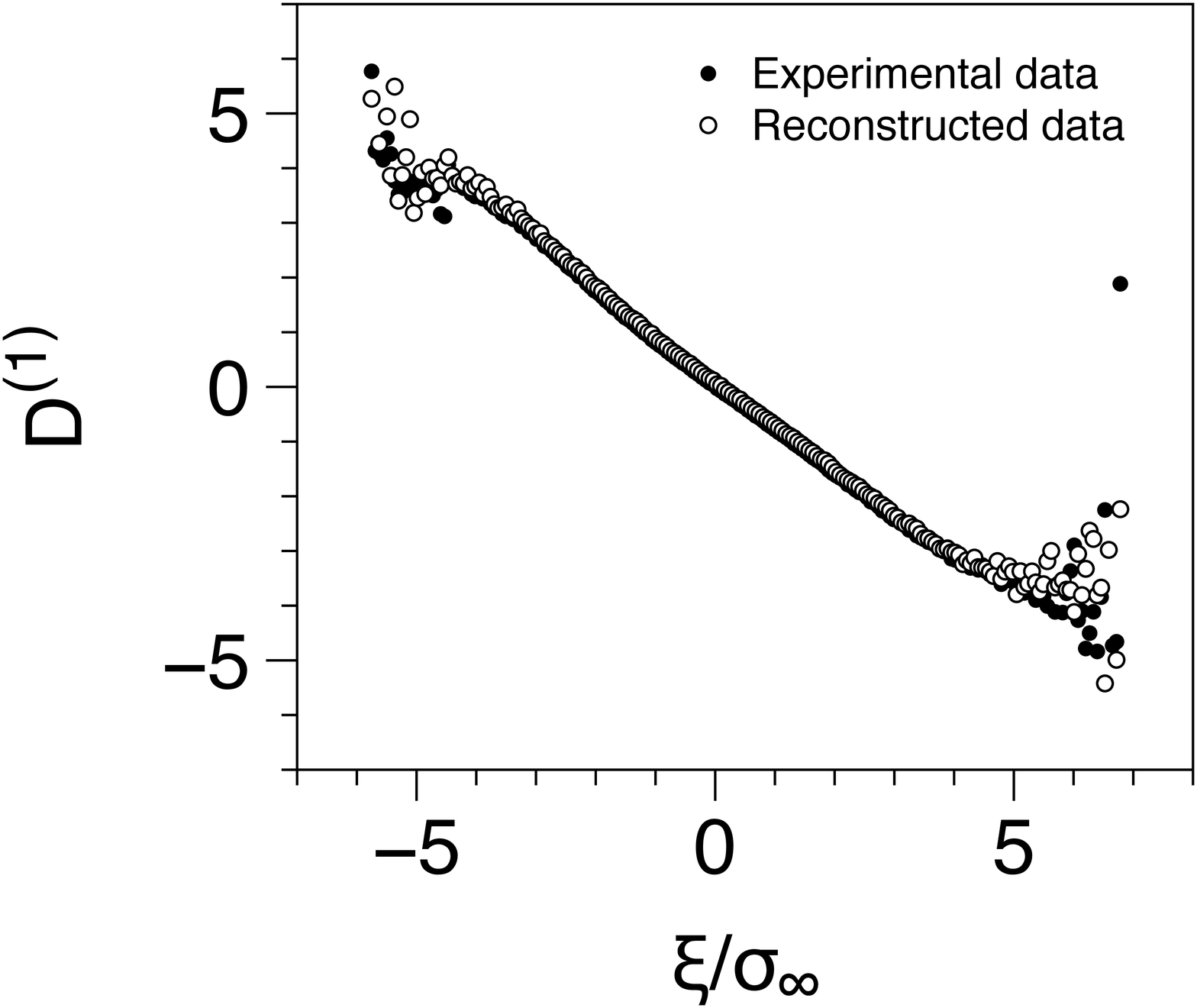}
        \includegraphics[width=.47\textwidth]{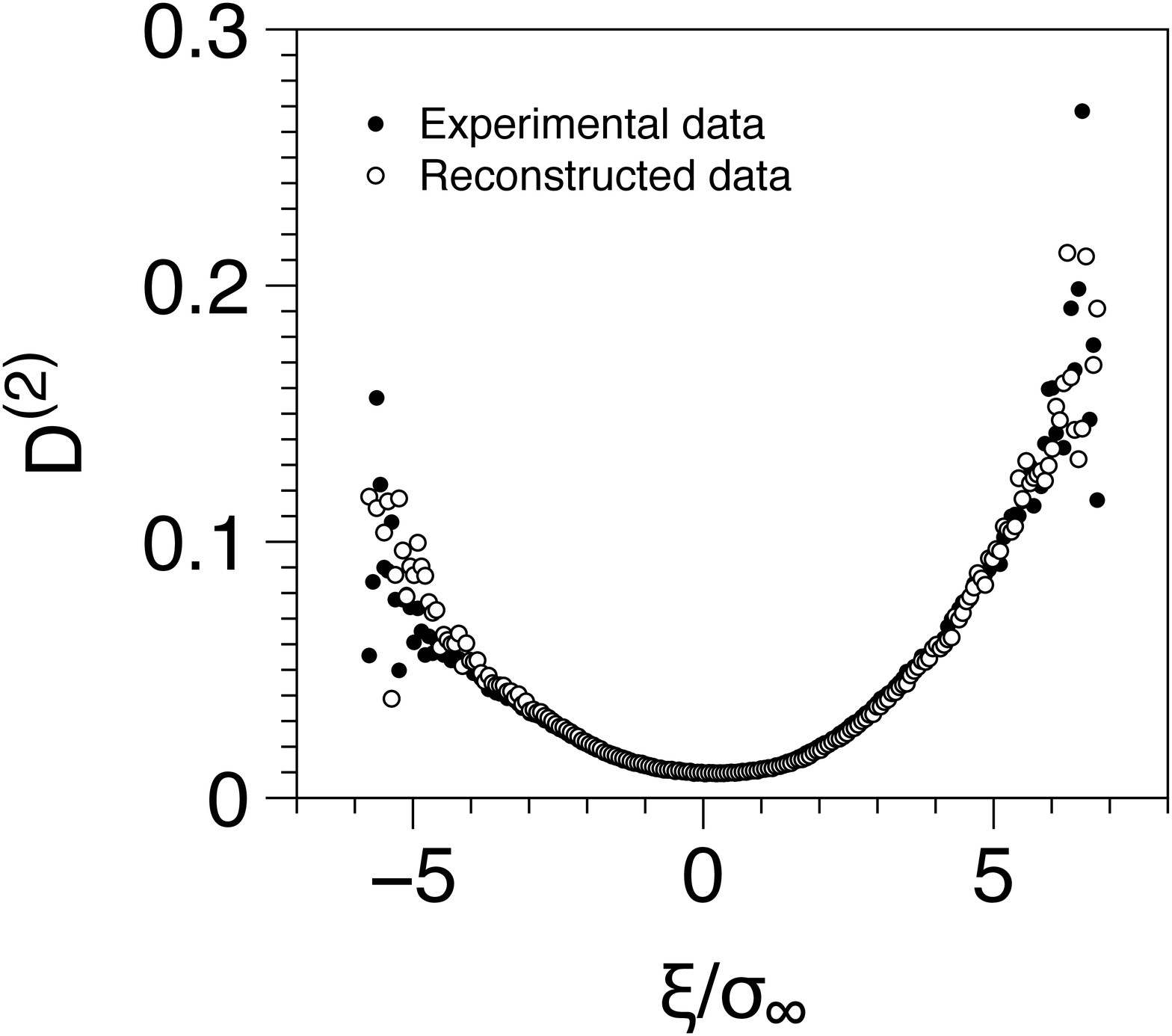}
  \end{center}
 \caption{the drift $D^{(1)}(\xi,\tau,h)$ and diffusion $D^{(2)}(\xi,\tau,h)$ coefficients of ocean wave data in time scale $\tau=280$ $s$. The black dots are the original data and hollow circles are from reconstructed time series.}
 \label{d1d2_recon}
\end{figure}

\begin{figure}[thb]
  \begin{center}
    \includegraphics[width=.47\textwidth]{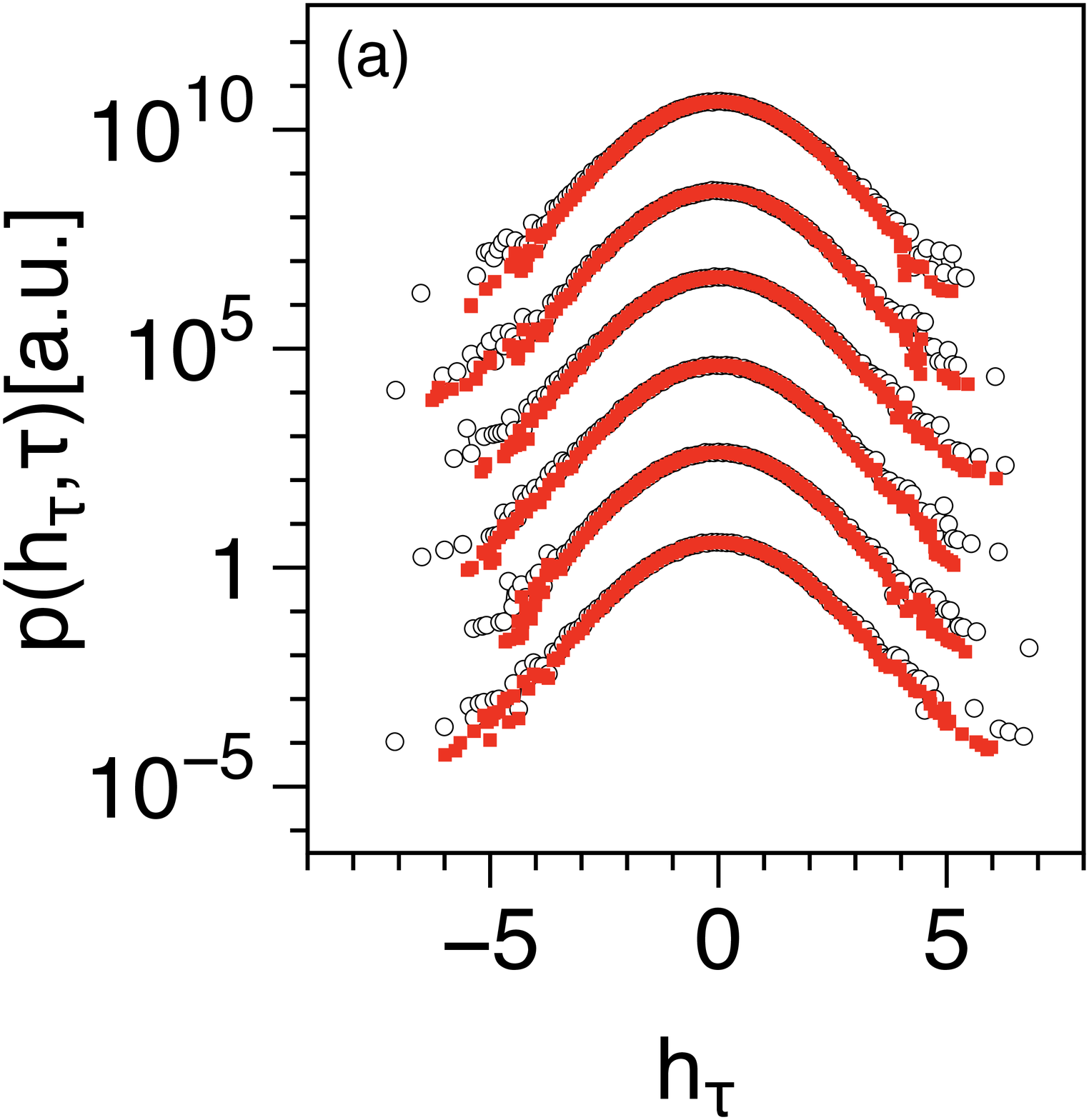}
        \includegraphics[width=.47\textwidth]{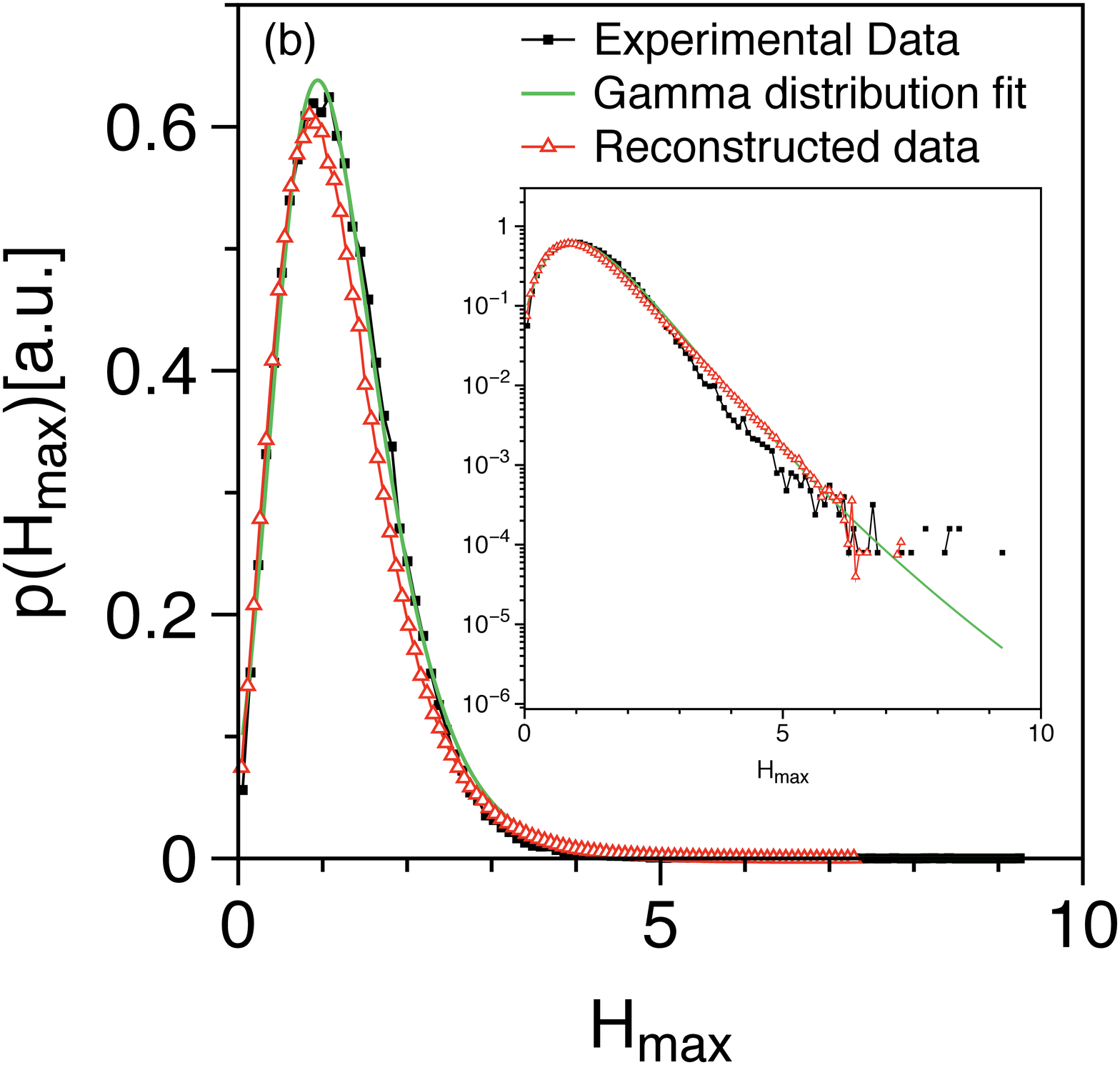}
  \end{center}
 \caption{(a) Empirical (hollow circles) and reconstructed (red filled symbols) PDFs for different scales. Time scales $\tau=14,28,42,56,70$ and $84 $ are chosen and PDFs are shifted in vertical direction for clarity of presentation. (b) Distribution of wave height maxima for empirical and reconstructed data in normal and log (inset plot) scale, which follows gamma distribution.   }
 \label{reconpdf}
\end{figure}

  %%%%%%%%%%%% %%%%%%%%%%%%%%%%%%%%%%%%%%%%%%%%%%%%%%%%%%%%%%%%%%%%%%%%%%%%%%%%%  
To illustrate that the reconstructed time series are indeed statistically similar to the measured wave data
we repeat the above mentioned Fokker-Planck analysis. In Fig. \ref{d1d2_recon} we show that from the surrogate data we obtain the same drift and diffusion coefficients.  Also the corresponding PDFs  $p(\xi_i, \tau_i)$   obtained from the measured data and from the numerical solution of the Fokker-Planck equation using the estimated drift and diffusion terms are the same as shown in Fig. \ref{reconpdf}(a). Furthermore the statistics of the wave height maxima are well grasped by the reconstructed data, see  Fig. \ref{reconpdf}(b). Both empirical and reconstructed data follow a generalized gamma distribution very well, as expected from \cite{Hadjihosseini2014}. From this verification of the obtained stochastic process we conclude that both, the empirical data and the reconstructed data, have the same multi-point statistics.
 
%%%%%%%%%%%% %%%%%%%%%%%%%%%%%%%%%%%         FIG 6           %%%%%%%%%%%%%%%%%%%%%%%%%%%%%%%%%  
%%%%%%%%%%%% %%%%%%%%%%%%%%%%%%%%%%%%%%%%%%%%%%%%%%%%%%%%%%%%%%%%%%%%%%%%%%%%

%%%%%%%%%%%% %%%%%%%%%%%%%%%%%%%%%%%         FIG 7         %%%%%%%%%%%%%%%%%%%%%%%%%%%%%%%%%

\begin{figure}[thb]  
 \setlength{\unitlength}{0.1\textwidth}
\begin{picture}(2,6)
\put(0.5,2){\includegraphics[width=0.48\textwidth]{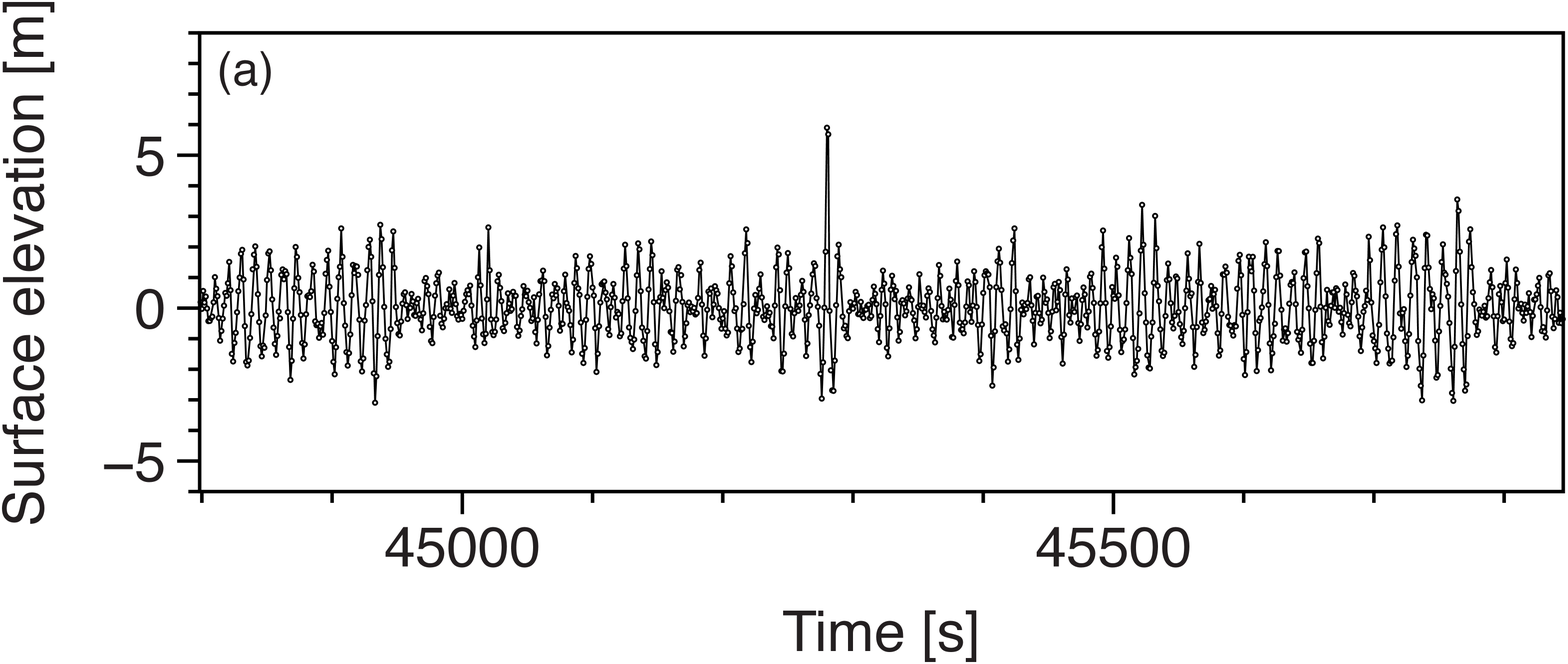}}
\put(5.3,4){\includegraphics[width=0.48\textwidth]{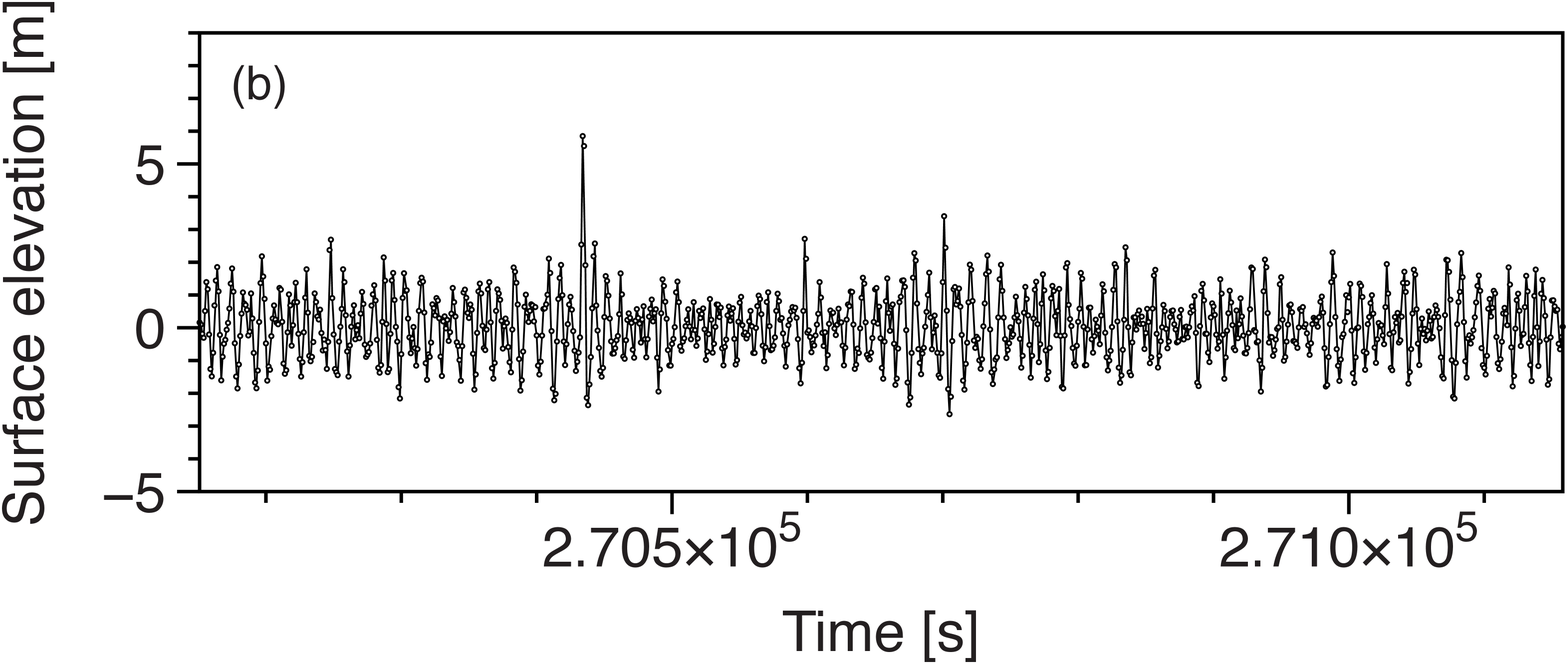}}
\put(5.3,2){\includegraphics[width=0.48\textwidth]{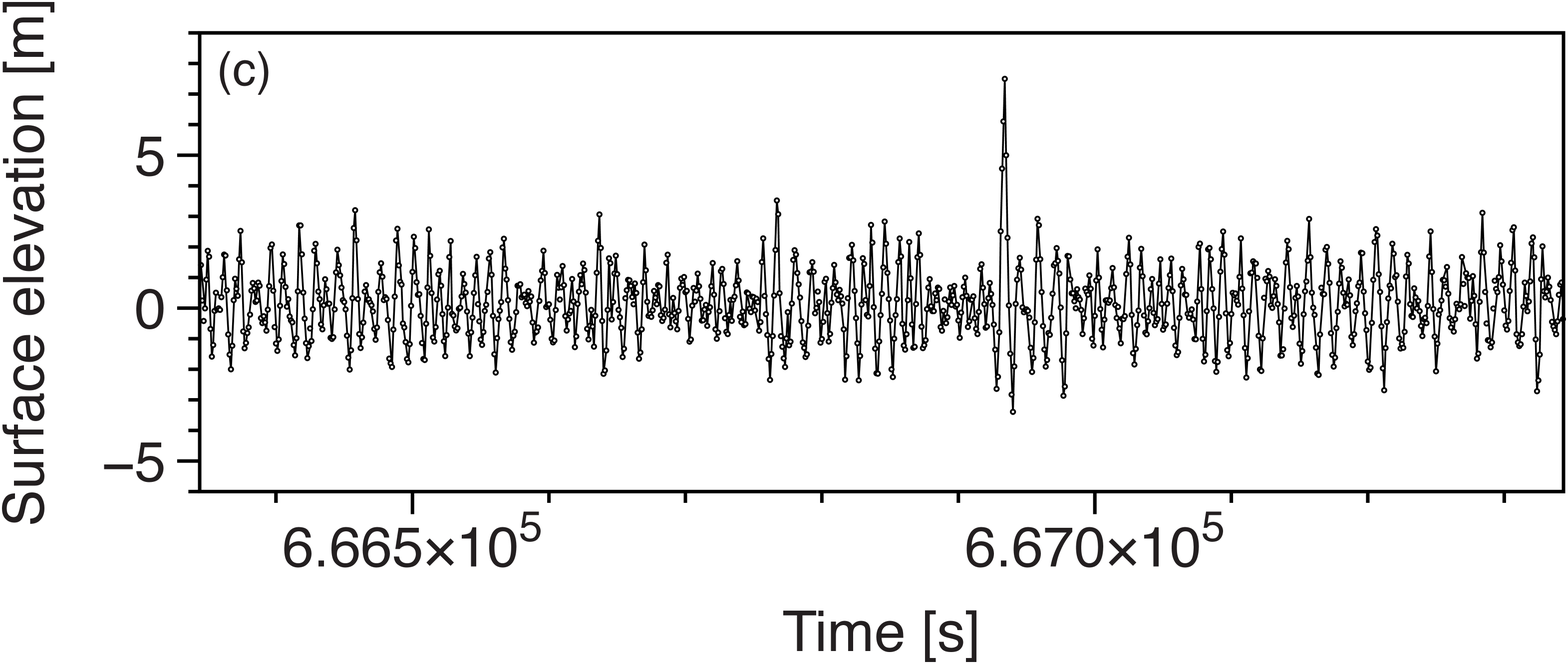}}
\put(5.25,0){ \includegraphics[width=0.48\textwidth]{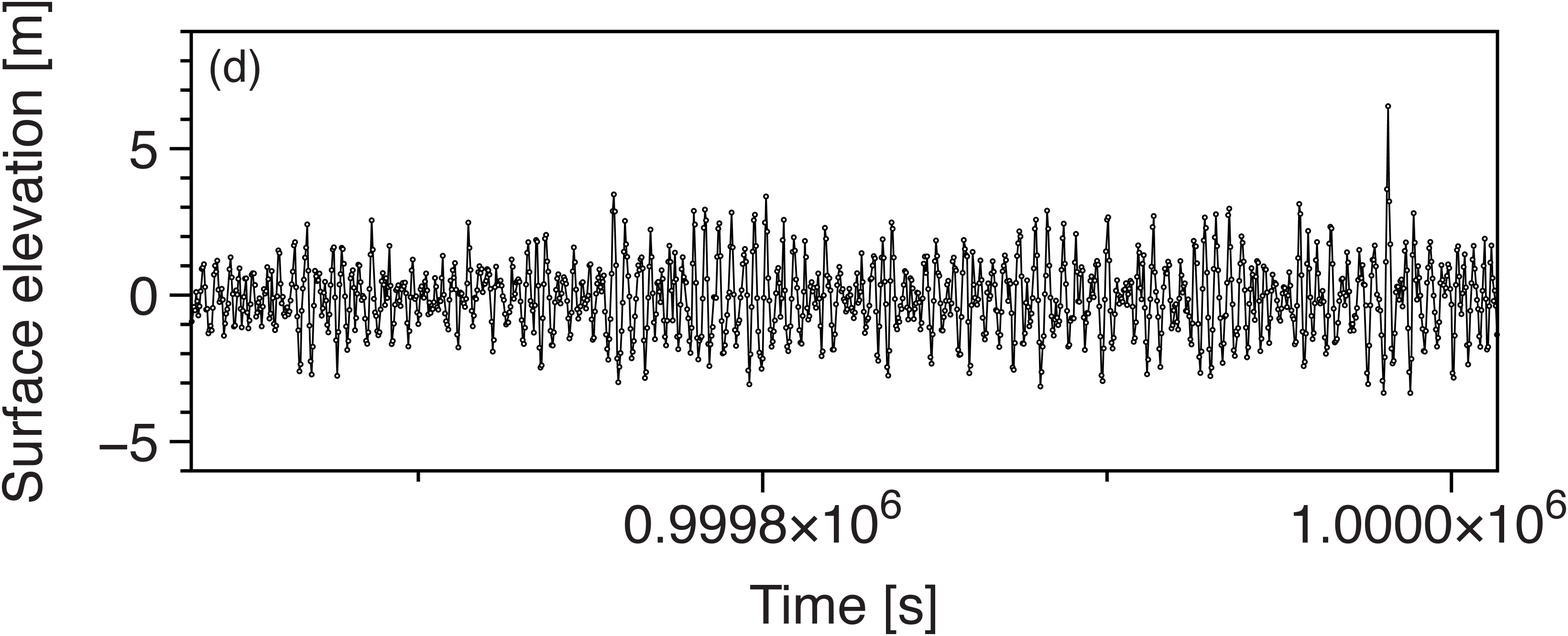}}
\end{picture}
 \caption{Three different part of  stochastic reconstructed time series (b,c and d) based on multi point PDFs, from the empirical data as initial conditions (a).}
     \label{rec}
 \end{figure}

Based on the proposed reconstruction of time series it is now possible to generate long synthetic time series
to work out further statistical features of the wave data.  We have chosen $1000$ data points of empirical data as initial condition and run it to produce $1.1\times 10^6$ synthetic data with sampling rate of $1$ $Hz$. In this reconstructed time series we have captured three events that we could consider as rogue waves, using the usual definition \cite{Akhmediev2010}, saying $h^*$ must be larger than 2 times the significant wave height, which is $2.4$ $m$ for our data. The corresponding three sections of the reconstructed time series are shown in Fig. \ref{rec} (b,c and d). Also, we performed $4096$ different runs of $2048$ seconds blocks. From these data  we captured $33$ time series with extreme values and a corresponding  waiting time of about $2.5\times10^5$ seconds to obtain an extreme event, or a rogue wave.

  %%%%%%%%%%%% %%%%%%%%%%%%%%%%%%%%%%%         FIG 8           %%%%%%%%%%%%%%%%%%%%%%%%%%%%%%%%%

\begin{figure}[thb]
  \begin{center}
    \includegraphics[width=.9\textwidth]{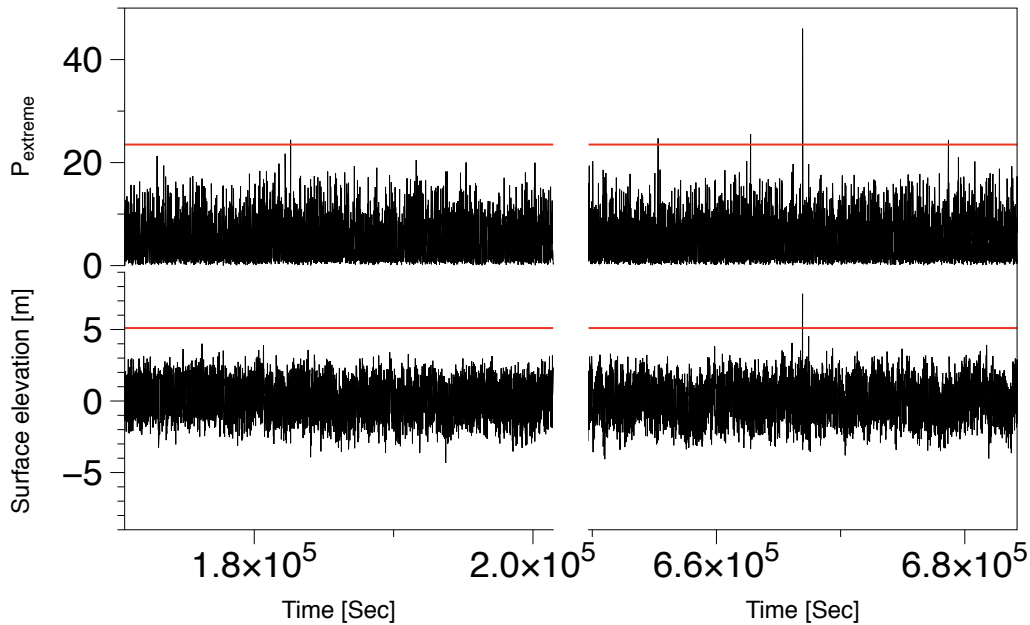}
    \end{center}
 \caption{Reconstructed time series (down) and probability of having an extreme event for each reconstructed point, $P_{extreme}$. (up) }
 \label{threshold}
\end{figure}

%%%%%%%%%%%% %%%%%%%%%%%%%%%%%%%%%%%%%%%%%%%%%%%%%%%%%%%%%%%%%%%%%%%%%%%%%%%%%  
Next we discuss the possibility to forecast emerging rogue waves. From the conditional probability, Fig. \ref{recon_time_series}(e) (Black curve) we can quantify the likelihood of the appearance of the measured amplitude of $h_r>5.2$ $m$  by integration 
%%%%%%%%%%%% %%%%%%%%%%%%%%%%%%%%       Equation  15     %%%%%%%%%%%%%%%%%%%%%%%%%%%%%%%%%%%%  

\begin{equation}
P_{extreme}=\int_{h_r}^{\infty}p(h^*|h_1,\tau_1; \dots; h_N,\tau_N) dh^*
\label{extreme_pdf}
\end{equation}
 %%%%%%%%%%%%%%%%%%  
and obtain 23.6\%. This likelihood $P_{extreme}(h_r > 5.2 \; m)$ can be evaluated for each time step and result in the
changing risk of emerging rogue waves, as shown in Fig. \ref{threshold}. This probabilistic characterization of extreme events 
returns some false alarms and as well some true hits.

%%%%%%%%%%%% %%%%%%%%%%%%%%%%%%%%%%%         FIG 9           %%%%%%%%%%%%%%%%%%%%%%%%%%%%%%%%%  
\begin{figure}[H]
  \begin{center}
    \includegraphics[width=.9\textwidth]{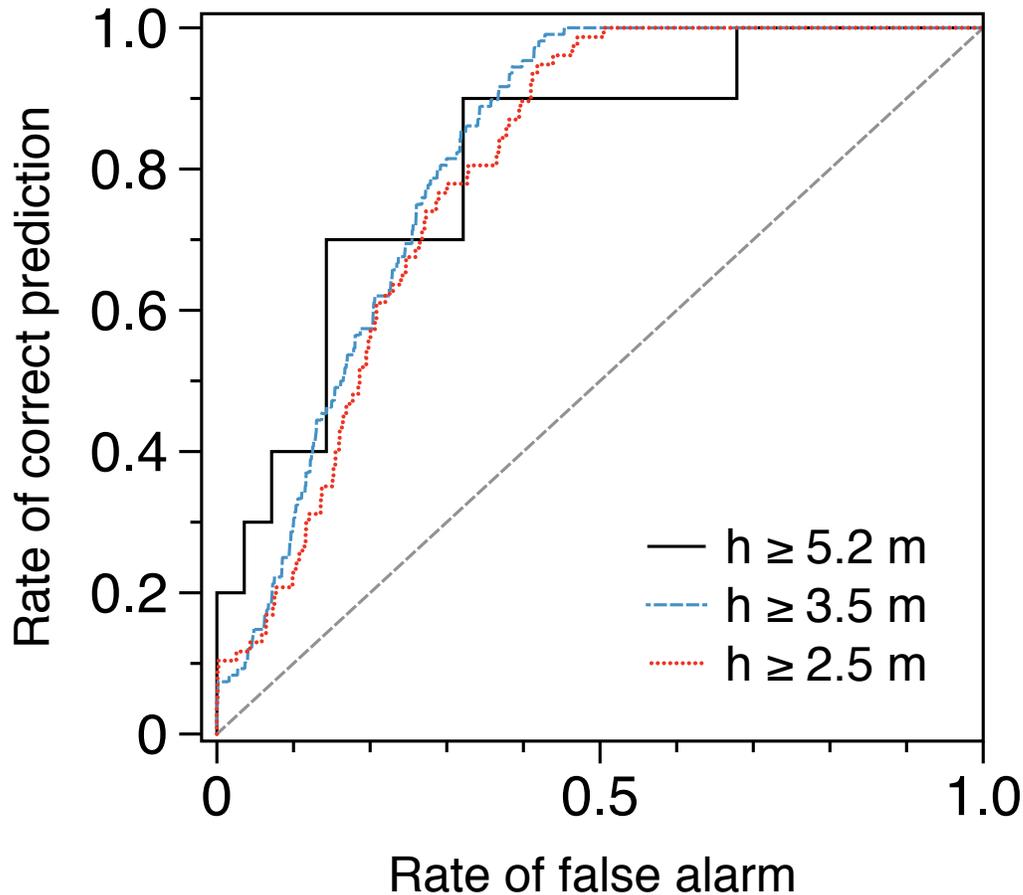}
    \end{center}
 \caption{ROC curve for three different estimation of $P_{exterme}$, for $h_r=5.2$ $m$ (solid black line), $h_r=3.5$ $m$ (dash blue line) and $h_r=2.5$ $m$ (dotted red line)  }
 \label{roc}
\end{figure}

%%%%%%%% %%%%%%%%%%%%%%%%%%%%%%%%%%%%%%%%%%%%%%%%%%%%%%%%%%%%%%%%%%%%%%%%%%%
 A common method to test the quality of a prediction is the receiver operating characteristic curve (ROC) \cite{Hanley1982,Kantz2006,Hallerberg2008}. The idea of ROC consists of comparing the rate of true predicted events with the rate of false alarm. The most quantitative index describing a ROC curve is the area under it which is known as accuracy. In Fig. \ref{roc} we have plotted ROC curves for our prediction, $P_{extreme}$, first by considering $h_r=5.2$ $m$ to detect the extreme event alarms. The corresponding ROC curve is plotted in black (solid) line. To investigate the robustness of our reconstruction method, we considered lower amplitude wave height for $h_r=2.5$ $m$ and $h_r=3.5$ $m$ and the corresponding ROC curves are shown in Fig. \ref{roc} in  red (dotted) and blue (dash) lines, respectively.  In all three cases we have accuracy of ROC curves bigger than $80\%$ which indicate that our multi-point procedure is a proper method for time series reconstruction and can be used for short time prediction purposes. 

%%%%%%%%%%%% %%%%%%%%%%%%%%%%%%%%%%%         FIG 10           %%%%%%%%%%%%%%%%%%%%%%%%%%%%%%%%%  
%%%%%%%% %%%%%%%%%%%%%%%%%%%%%%%%%%%%%%%%%%%%%%%%%%%%%%%%%%%%%%%%%%%%%%%%%%%

\section{Conclusions}

We have presented a new approach for a comprehensive analysis of the complexity of ocean wave dynamics. The complexity of multi-point statistics can be simplified by a three point closure, based on which an arbitrary N-Point statistics can be expressed by a hierarchy of nested three point statistics ordered in a cascade like structure. We have been able to show for the first time that by our stochastic approach not only the joint N-point statistics can be grasped but also extreme events, rogue waves, can be captured statistically. We have also shown how for each instant in time the conditional probability of the next wave height can be determined. As the height profile of waves changes from moment to moment, also the probability of the next value of the wave height is changing dynamically. These changes may thus clearly give rise to measures indicating the risk of the appearance of rogue waves ahead of their actual emergence. Most interestingly this was possible although in the measured data only one event of rogue wave was recorded. From our analysis of the occurrence probabilities it becomes clear that the rogue wave for this wave conditions is integral part of the entire complex stochastic. In our opinion this can only be achieved as we were able to  crave out an N-point approach for this complex system.

\section*{Acknowledgement}
Ali Hadjihosseini wish to express his gratitude to David Bastine, Pedro Lind and Alexey Slunyaev for helpful discussions. This work was supported by VolkswagenStiftung, grant number 88482.

%%%%%%%%%%%%%%%%%%%%%%%%%%% BIBLIOGRAPHY  %%%%%%%%%%%%%%%%%%%%%%%%%%%%%%%%%
\section*{Refrences}
\bibliographystyle{unsrt}
\bibliography{mybib}
\end{document}